# Electron Confinement-Induced Plasmonic Breakdown in Metals

Prasanna Das[1], Sourav Rudra[1], Dheemahi Rao[1], Souvik Banerjee[1], Ashalatha Indiradevi Kamalasanan Pillai[2], Magnus Garbrecht[2], Alexandra Boltasseva[3], Igor V. Bondarev[4], Vladimir M. Shalaev[3], and Bivas Saha[1,5]*

[1]*Chemistry and Physics of Materials Unit and International Centre for Materials Science, Jawaharlal Nehru Centre for Advanced Scientific Research, Bangalore 560064, India*

[2]*Sydney Microscopy and Microanalysis, The University of Sydney, Camperdown, NSW 2006, Australia*

[3]*Elmore Family School of Electrical and Computer Engineering, Purdue Quantum Science and Engineering Institute, and Birck Nanotechnology Center, Purdue University, West Lafayette, IN 47907, USA*

[4]*Department of Mathematics & Physics, North Carolina Central University, Durham, NC 27707, USA*

[5]*School of Advanced Materials and Sheikh Saqr Laboratory, Jawaharlal Nehru Centre for Advanced Scientific Research, Bangalore 560064, India*

*Correspondence emails: bsaha@jncasr.ac.in and bivas.mat@gmail.com

**Fully referenced summary paragraph:**

Plasmon resonance in metals represents the collective oscillation of the free electron gas density and enables enhanced light-matter interactions in nanoscale dimensions [1–4]. Traditionally, the classical Drude model describes the plasmonic excitation, wherein the plasma frequency exhibits no spatial dispersion [5]. Here, we show conclusive experimental evidence of the breakdown of the plasmon resonance and a consequent photonic metal-insulator transition in an ultrathin archetypal refractory plasmonic material, hafnium nitride (HfN). Epitaxial HfN thick films exhibit a low-loss and high-quality Drude-like plasmon resonance in the visible spectral range [6,7]. However, as the film thickness is reduced to nanoscale dimensions, the Coulomb interaction among electrons increases due to the electron confinement, leading to the spatial dispersion of the plasma frequency [8,9]. Importantly, with the further decrease in thickness, electrons lose their ability to shield the incident electric field, turning the medium into a dielectric. The breakdown of the plasmon resonance in epitaxial ultrathin metals could be useful for fundamental physics studies in transdimensional regimes [10] and novel photonic device applications.

**Main**

Plasmons are quantized quasiparticles of collective free-electron oscillation in conductive media, enabling light confinement at nanoscale dimensions [1–4]. Plasmon resonance leads to intense optical absorptions and local field enhancement in nanostructures [11] and is used extensively for various applications such as photocatalysis [12], local heating [13], photovoltaics [14], sensing [15], microscopy [16], optical communications [17], and nonlinear optics [18]. Traditionally, the classical Drude model is used to describe plasmon resonance, wherein the screened plasma frequency ($\omega_p$) depends on the carrier concentration ($n$), effective mass ($m^*$), and the core dielectric constant ($\varepsilon_{core}$) of the medium [5]. The Drude model has been used extensively to describe the plasmonic properties of metallic films and nanoparticles [19,20], doped semiconductors [21], layered materials such as graphene [22], and transition metal dichalcogenides in the visible-to-infrared spectral ranges [23]. However, as $\omega_p$ does not exhibit any spatial dispersion (thickness-dependence) according to the Drude model, the tunability of plasmon resonance can be achieved only with the changes in the doping concentration of semiconductors in the infrared (IR) spectral ranges [21]. For Drude metals exhibiting plasmon resonances in the visible spectral ranges, tuning the plasmon resonance is exceptionally challenging as the carrier concentrations cannot be varied readily.

However, a recent non-local Drude dielectric response theory based on the Keldysh-Rytova (KR) potential that incorporates electron confinement in a thin plasmonic medium predicts a spatial dispersion of the plasmon resonance and increased dissipative loss at the $\omega_p$ (Fig. 1a and 1b) [8,9]. The theory further suggests that in the limit where the dielectric constant ($\varepsilon^{core}$) of the medium is greater than the surroundings (substrate ($\varepsilon_1^{core}$) and superstate ($\varepsilon_2^{core}$)) and in ultrathin films (see supplementary information (SI) for details), the strong unscreened in-plane Coulomb potential confines carriers. Due to such electronic confinement, plasmonic metals lose their ability to respond to an incident electric field, turning effectively into a dielectric. Such breakdown of the plasmon oscillations in metals is analogous to the electric field-induced dielectric breakdown in insulators and could drastically alter plasmonic device performance [24]. The momentum-averaged density of electronic states also qualitatively explains the plasmonic breakdown and the consequent electronic metal-insulator transition (MIT) (Fig. 1c). In the thick plasmonic films, the electron states are delocalized and the Fermi energy ($E_F$) lies inside the conduction band, leading to their Drude-like metallic response. However, as the film thickness is reduced, the increased Coulomb interaction among charges localizes their states close to the $E_F$, leading first to transdimensional (TD) plasmonic medium [25] where the mobility edge ($E_M$, the energy that separates localized and non-localized states) is below the $E_F$ and eventually to the plasmon resonance breakdown when more electron states are localized and the $E_M$ shifts above the $E_F$. Due to the plasmonic breakdown, the material effectively becomes a dielectric, termed as KR insulator here.

The possibility of observing the electron confinement-induced plasmonic breakdown in metals is intriguing and could provide insights into the physical phenomena associated with strongly correlated systems in TD materials as well as means to control it. Thus far, experimental observation of the MIT in TD materials is challenging due to the materials' limitations. Traditionally, noble metals such as gold (Au) and silver (Au) are regarded as the best-known plasmonic materials for the visible spectral range. However, depositing ultra-thin films of Au, Ag, or other noble metals is challenging as they exhibit island (Volmer-Weber) growths due to their large surface energies (1-2 J/m$^2$) [26]. Moreover, noble metals are not structurally and morphologically stable at elevated temperatures (1000 ℃) and are incompatible with the complementary metal oxide semiconductor (CMOS) process, limiting their practical applications [27]. On the other hand, transition metal nitrides (TMNs) such as TiN and HfN have emerged as an alternative plasmonic material to Au and Ag, respectively, and exhibit many interesting plasmonic properties [27,28]. TD epitaxial TiN films have been deposited recently on (001) MgO substrates and a spatial dependence of $\omega_p$ has been demonstrated [29]. However, the TD TiN films retain their metallic characteristics even at the smallest thickness ( 2 nm), thereby precluding the observation of the plasmonic breakdown [30]. Unambiguous demonstration of the plasmonic breakdown and the MIT in epitaxial ultrathin metals could unlock new electron confinement-related physical phenomena and enable novel optical and optoelectronic devices, especially in flat optics utilizing metasurfaces. In this work, we present the first conclusive experimental evidence of the electron confinement-induced plasmonic breakdown and a consequent photonic metal-to-insulator transition in an ultrathin refractory metal, hafnium nitride (HfN).

HfN is an ambient-stable, corrosion-resistant TMN with a melting temperature of ~3600K. HfN crystalizes in the rocksalt structure and exhibits a superconducting ground state with a transition temperature of ~ 5 – 10K [31]. HfN films exhibit a high electrical conductivity of ~ 3.7 × $10^4$ Sm$^{-1}$ and carrier concentration of ~ 6.6×$10^{22}$ cm$^{-3}$ [7]. A long photo-excited carrier lifetime [32], large phononic bandgap [33], and high mechanical hardness have made HfN attractive for hot-carrier solar cell and tribology applications.

HfN films deposited on (001) MgO substrates exhibit high-quality and low-loss Drude-like plasmon resonance in the 350-420 nm spectral range that is comparable to the optical properties of noble metals. With an optimized deposition condition, recent research has also shown that HfN films exhibit a high solar reflectivity of 90% and infrared reflectivity of 95%, which make it an alternative plasmonic material to Ag for solar mirror and IR reflector applications [6,7]. Interestingly, compared to TiN or Ag, the Drude damping coefficient ($\gamma_D$) of HfN is much higher due to stronger electron-phonon interactions, dislocations, grain boundary, and defect scattering (see Extended Data Table I, Section II, and Section V in SI), which make it a potential candidate for studying confinement-related physical phenomena.

**Confinement-induced photonic metal-insulator transition**

Epitaxial bulk (150 nm thick) and ultrathin (thickness from 2 nm to 15 nm) HfN films are deposited on (001) MgO substrates at 900℃ substrate temperature utilizing an ultrahigh vacuum magnetron sputtering (see SI for details). The dielectric permittivity of the deposited films is measured with spectroscopic ellipsometry in the 210-2500 nm spectral range (see SI for details). Results show that thick HfN film exhibits Drude-like plasmonic response with an epsilon-near-zero (ENZ) cross-over wavelength (where the real part of the dielectric permittivity ($\varepsilon'$) changes from positive to negative thus crossing zero) at 415 nm (Fig. 1d). The $\varepsilon'$ increases monotonically and exhibits excellent metallic response, represented by high $|\varepsilon'|$ at long wavelengths. Similarly, the imaginary component of the dielectric permittivity ($\varepsilon''$) or material optical loss increases at long wavelengths due to free-electron Drude absorption. Low optical loss near the ENZ wavelength ($\lambda_{ENZ}$) leads to the high plasmonic *figure-of-merit* [7], long propagation length, and large decay length in HfN, as shown in Section I of SI.

However, as the thickness of the film is reduced, $|\varepsilon'|$ decreases progressively in the visible and near-IR spectral ranges (Fig. 1e), suggesting the reduced metallicity of the films. The ENZ wavelength exhibits a red shift from 478 nm in 15-nm-thick to 787 nm in 4-nm-thick films. Such spatial dispersion of the ENZ wavelength departs from the classical Drude metallic characteristics in HfN and arises due to electron confinement from enhanced interactions among charges in the TD films. When the thickness is reduced further (less than 4 nm), $\varepsilon'$ becomes positive over the entire spectral range, thus exhibiting the breakdown of the plasmon resonance and the onset of the photonic MIT. The cross-over from the TD-plasmonic to KR-insulating regime can be obtained from the non-local Drude dielectric response theory that takes into account confinement effects in ultrathin films (see Section II in SI for details) [8,34], where $\varepsilon'$ is expressed by the relation

$$\varepsilon^{core}\left(1 - \frac{(\omega_p^{3D})^2}{(\omega^2+\gamma_D^2)\left(1+\frac{\varepsilon_1^{core}+\varepsilon_2^{core}}{\varepsilon^{core}kd}\right)}\right) > 0 \qquad (1)$$

Here, $\omega_p^{3D}$, $\gamma_D$, and $k$ are the unscreened plasma frequency in 3D-bulk film, Drude damping coefficient, and the absolute value of the electron wavevector, respectively. With reduction in the thickness $d$, the repulsive KR potential among charges becomes independent of $\varepsilon^{core}$. In this case the second term in parenthesis of Eq. (1) becomes proportional to $(\omega_p^{3D})^2 \times kd$, which ultimately makes it less than one, and the material loses its ability to shield the incident electric field, completing its transition to the KR-insulator state [8].

Optical loss or $\varepsilon''$ in the TD-plasmonic and KR-insulating regime decreases with the reduced film thickness at long wavelengths (Fig. 1f). Such a decrease in $\varepsilon''$ originates from the reduced carrier concentration in thinner films (see Extended Data Table II). Interestingly, $\varepsilon''$ at ENZ wavelength increases substantially with decreasing film thickness from 150 nm to 4 nm due to increased $\gamma_D$ (Fig. 1g and Fig. 1h). Such an increase in $\gamma_D$ with the decreasing film thickness appears due to a reduction in the electron mean free path owing to different scattering mechanisms. In ultrathin films, the main contribution to the damping originates not only from the increased electron-electron interactions, but could also have contributions from surface roughness, electron-phonon, dislocations, and grain boundary scatterings. As electrons are confined progressively in thinner films their ability to respond to the incident electric fields decreases, which not only reduces $|\varepsilon'|$ in the TD-plasmonic regime but eventually leads to the MIT. The red shift in the ENZ wavelength or decrease in the $\omega_p$ in the TD-plasmonic regime is also associated with the vertical confinement of the electrons. Experimental results show a $\sqrt{d}$ dependence of $\omega_p(k)$ with the decreasing film thickness [29] which is consistent with the theoretical predictions (see Section II C in SI for detailed discussion) [8,34].

Experimental reflection and transmission measurements verify the transition first from a Drude metal to a TD-plasmonic medium and then subsequent KR-insulator optical transition in HfN films. Reflection (and transmission) spectra of 15-nm-thin HfN film show a clear dip (and peak in transmission) at ~ 440 nm, near to its ENZ wavelength (Fig. 1i and 1j). However, as the thickness of the film is reduced, the sharpness of the reflection dip (consequent transmission peak) becomes less prominent in the TD films (see Extended Data Fig. 1). The HfN film with 2 nm thickness exhibits a near constant ~ 98% transmission above ~550 nm that manifests its dielectric nature. Absorptions below 550 nm are associated with interband losses, as modelled using Lorentz oscillators. The optical images of HfN films (see inset in Fig. 1j) show the variation in transparency with changes in the film thickness.

**Temperature-dependent optical properties**

Temperature-dependent dielectric permittivity provides further evidence for the three different optical regimes in HfN films. We observe that the $|\varepsilon'|$ of a thick HfN film decreases with the increasing temperature from 200K to 700K while the $\varepsilon''$ increases due to the increased $\gamma_D$ at high temperatures (see Fig. 2a, Fig. 2b, and Section II H in SI). In the TD-plasmonic regime (represented by the 5-nm-thick film), $|\varepsilon'|$ also decreases with the increasing temperature due to an increase in $\gamma_D$ [35]. Compared to the Drude metallic regime of HfN, the $\varepsilon''$ is decreasing to a lesser degree at higher temperatures due to a near-constant $\omega_p$. Such invariance of $\omega_p$ with the temperature is a unique feature of the TD-plasmonic medium and was earlier predicted theoretically (see Section II C in SI for details) [36]. Our finding is the first experimental verification of this behavior where both 10-nm-thick and 4-nm-thick TD-plasmonic HfN films also show near-constant $\omega_p$ over the entire wavelength range (see Fig. 2c). In the KR-insulating regime (represented by the 2 nm film), $\varepsilon'$ is positive throughout the temperature range and shows little variations. On the other hand, similar to the TD-plasmonic regime, the $\varepsilon''$ reduces slightly at higher temperatures.

**Electronic metal-insulator transition**

Thickness and temperature-dependent electronic transport measurements further correlate with the optical properties. Thick HfN film exhibits a low room temperature resistivity of $6.4 \times 10^{-5}$ Ohm-cm consistent with its metallic nature. However, as the thickness is reduced from 150 nm to 4 nm, resistivity increases by nearly five times in the TD-plasmonic regime at room temperature (see Fig. 3a). Interestingly, with a further reduction in thickness, resistivity increases drastically by more than three orders of magnitude in thinner films (< 4 nm). Such a drastic increase in resistivity suggests the onset of electronic confinement and a metal-insulator electronic transition, which correlates with the plasmonic breakdown in the ultrathin films. These results further suggests that in the KR-insulating regime, large in-plane electronic interactions and the two-dimensional nature of the potential localize all states and $E_F$ is below $E_M$ as shown in Fig. 1c.

Consistent with resistivity, the carrier concentration (measured with the Hall experiment, see Extended Data Table II) also decreases progressively in the TD-plasmonic regime but shows nearly two orders of magnitude decrease in ultrathin films (Fig. 3b). Within the Drude-metallic and TD-plasmonic regime, the variations of resistivity with thickness have been fitted with the DC-optical conductivity, considering modified dielectric permittivity (see Section II F in SI) [37]. As the insets show, the theoretical model matches the experiment results well. Note that even the most resistive film in the KR-insulating regime, 2 nm HfN, exhibits a resistivity of $2.41 \times 10^{-1}$ Ohm-cm (and carrier concentration of $3.70 \times 10^{20}$ cm$^{-3}$) that is significantly lower in magnitude compared to the resistivity of other well-established band insulators such as SiO$_2$, AlN, etc. [38,39].

Temperature-dependent (50 K-400 K) resistivity provides further evidence about the ensuring electronic and optical metal-to-insulator phase transition in HfN (see Fig. 3c). Thick 150 nm film, representing the Drude-metallic regime, shows a positive temperature coefficient of resistivity (TCR) typical for a good metal. The measured TCR of $0.9 \times 10^{-3}$/°C in HfN matches well with that of Au, Ag, Cu, and other metals (see Table S6 in SI). As the thickness is reduced, initially, the films retain their positive TCR, such as in 10 nm film. However, the sign of TCR changes starting from 5-nm-thick film that exhibits a slight increase in the resistivity at low temperatures, highlighting the onset of electronic confinement. As the thickness is reduced further, the strength of confinement increases drastically for the 2.5 nm and 2 nm films as indicated by a significant increase in the resistivity at low temperatures and the nearly unchanged magnetoresistance (see SI, Fig. S5). The experimental resistivity-vs.-temperature curves in thinner films are fitted with the nearest neighbour hopping (NNH) and Mott variable range hopping model (MVRH) [40]. Although the physical origin behind the localization of carriers is different in MVRH and the present case, the Mott temperature ($T_M$) provides a measure of the strength of confinement in the films (see Section III in SI). These results clearly show that with shrinking thickness, HfN thin films undergo a metallic-to-insulating electronic transition with a colossal eight orders of magnitude increase in the resistivity at low temperatures in 2-nm-thick HfN film compared to 150-nm-thick film representing the bulk limit. Therefore, the electrical measurements reconcile the optical properties measured above and provide evidence for the breakdown of plasmon oscillations in HfN films.

**Promise for Wigner Crystallization**

Furthermore, utilizing the repulsive KR interaction potential as the electrostatic interaction energy and the mean electron kinetic energy per particle as $\langle E_{kin} \rangle = \hbar^2 \pi N_{2D}/(2m^*)$, where $N_{2D} = N_{3D}d = 1/(\pi \bar{\rho}^2)$ is the surface charge density, the dimensionless Platzman-Fukuyama (PF) ratio [41] of potential interaction energy and mean kinetic energy representing confinement strength is calculated as,

$$\Gamma_0(\tilde{d}, n) = \frac{1}{\bar{\varepsilon}\tilde{\varepsilon}\tilde{d}n}\left\{H_0\left(\frac{1}{\tilde{\varepsilon}\tilde{d}\sqrt{\pi n}}\right) - Y_0\left(\frac{1}{\tilde{\varepsilon}\tilde{d}\sqrt{\pi n}}\right)\right\}. \qquad (2)$$

where, $\bar{\varepsilon} = (\varepsilon_1^{core} + \varepsilon_2^{core})/2$, $\tilde{d} = d/a_B$ is the dimensionless thickness, $n = N_{2D}a_B^2 = N_{3D}da_B^2$ is the dimensionless surface density, $H_0$ and $Y_0$ are the Strove and Neumann functions, respectively, and $a_B = 0.529/2$ Å is the Bohr radius of the 2D hydrogen atom to set up the in-plane distance scale. Calculated $\Gamma_0(n,d)$ (see Table S8) increases drastically as the thickness of the films and the corresponding carrier concentration decreases rapidly (Fig. 4a), which suggests a strong electron confinement in the TD plasmonic and KR insulator regimes. Interestingly, for the thinnest 2 nm HfN, $\Gamma_0(n,d)$ exceeds 1 satisfying the well-established PF condition for Wigner crystallization [41]. These calculations suggest the exciting possibility of achieving Wigner crystallization (*40, 41)* in TD materials (see Section IV in SI for details), however, its unambiguous experimental verification remains as a future avenue for research. Moreover, achieving $\Gamma_0(n,d)$ above 1 opens interesting opportunities to study strongly correlated regimes in TD materials where several key parameters such as the electron density, the potential interaction energy, and the average kinetic energy can be tailored simply by changing the material thickness and dielectric environment.

**First-principles calculations**

First-principles density functional theory (DFT) calculations are used further to explain the role of electronic structure, phonons, and electron-phonon coupling on the optical responses of HfN and to model the dielectric response. The electronic structure of HfN shows that the valence band primarily comprises of N-*2p* states similar to other TMNs such as ScN [42,43], TiN [44], and CrN [45], while the conduction band exhibits Hf-*5d* characteristics (see Fig. 4b). The peak in $\varepsilon''$ at 5.63 eV in practically all HfN films corresponds to the inter-band transition between N 2p-Hf 5d, valence band to the unoccupied conduction band at the Γ point, as shown in Fig. 1f. Though the conventional DFT calculations cannot incorporate the electron confinement effects in thinner films [46,47], a comparison of the electronic densities of states in bulk and a bilayer of HfN provides significant insight. As presented in Fig. 4c, the valence bandwidth and distance of the valence band tail from $E_F$ are lower in thinner films than in the bulk HfN. Since localization generally sets in from the band tails, the narrower bands close to the $E_F$ facilitate the electronic confinement [48]. Additionally, time-dependent (TD)-DFPT calculations that model the dielectric constants show that the experimental $\varepsilon'$ and $\varepsilon''$ can be well-captured for the bulk HfN with a low concentration of Hf vacancies (see Section V in SI ) [49].

**Structural Characterizations**

Finally, high-resolution X-ray diffraction (HRXRD) and transmission electron microscopy (HRTEM) imaging evidence the structural quality of the HfN films. Fig. 4d shows a symmetric $2\theta - \omega$ HRXRD 002 peak of thick HfN representing a lattice constant of 4.52 Å, consistent with its bulk values [50]. Though the 002 HRXRD peaks are broader in thinner films due to the presence of residual strain, they maintain their layer coherency and epitaxy as evidenced in the pole figure mapping (Fig. 4e and Extended Data Fig. 2) as well as X-ray reflectivity (XRR) analysis (Fig. 4f). High-angle annular dark field-scanning transmission electron microscopy (HAADF-STEM) images (see Extended Data Fig. 3) show cube-on-cube growth of (002) HfN on (001) MgO substrates with an epitaxial relationship [001] (001) HfN ∥ [001] (001) MgO substrates. Further details on the structural characterizations are presented in the SI.

In summary, we present the first conclusive experimental evidence of electron confinement-induced plasmonic breakdown and a photonic metal-insulator transition in an archetypal ultrathin refractory plasmonic metal, HfN. The strong in-plane Coulomb interaction among charges in transdimensional HfN films leads to a spatial dispersion of its plasma frequency and higher Drude damping coefficient. Eventually, electrons lose their ability to screen the incident electronic field, making the material a dielectric. Non-local dielectric response theory based on the Keldysh-Rytova (KR) potential explains the experimental observations. The breakdown of plasmon resonance in ultrathin HfN is akin to the dielectric breakdown in insulators and could enable novel active metasurface and metamaterial concepts where electron confinement is used to tune the optical properties of constituent plasmonic materials. Moreover, this approach could be used to study new physics of strongly correlated regimes in TD materials.

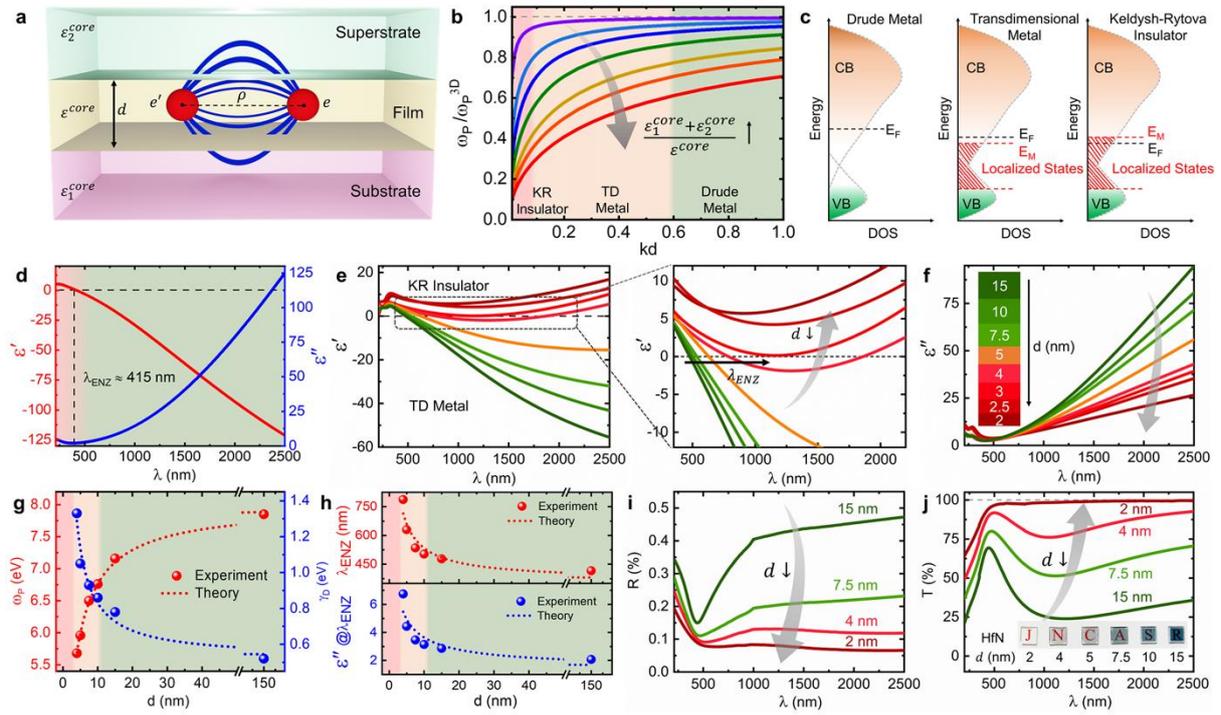

**Fig. 1 | Electron confinement-induced plasmonic breakdown in HfN.** (**a**) Schematic of a plasmonic metal sandwiched between a substrate and a superstate. As the film thickness (*d*) decreases, interaction potential among charges acquires an in-plane character and increases drastically, leading to electron confinement. (**b**) Due to the increased confinement, the plasma frequency (theoretically calculated) in transdimensional (TD) metals acquires a spatial dispersion and decreases with the reduced film thickness [8]. (**c**) Schematic representing the localization of the electronic states due to the electron confinement. (**d**) The real ($\varepsilon'$) and imaginary ($\varepsilon''$) components of the dielectric permittivity of a thick HfN film. (**e**) The $\varepsilon'$ of HfN films as a function of thickness. The inset shows a clear redshift of the epsilon-near-zero (ENZ) wavelength ($\lambda_{ENZ}$) in the TD films. (**f**) The $\varepsilon''$ decreases with the decreasing film thickness at long wavelengths. (**g**) For the TD films, the plasma frequency ($\omega_p$) and the Drude damping coefficient ($\gamma_D$) exhibit a spatial dispersion. (**h**) Variation of $\lambda_{ENZ}$ and optical loss ($\varepsilon''$) at the ENZ wavelength (experimental and theoretical) with the film thickness. Thickness-dependent (**i**) reflection and (**j**) transmission spectra of HfN films. The transmission of films with thickness from 2 nm to 15 nm are represented with the "JNCASR" written below.

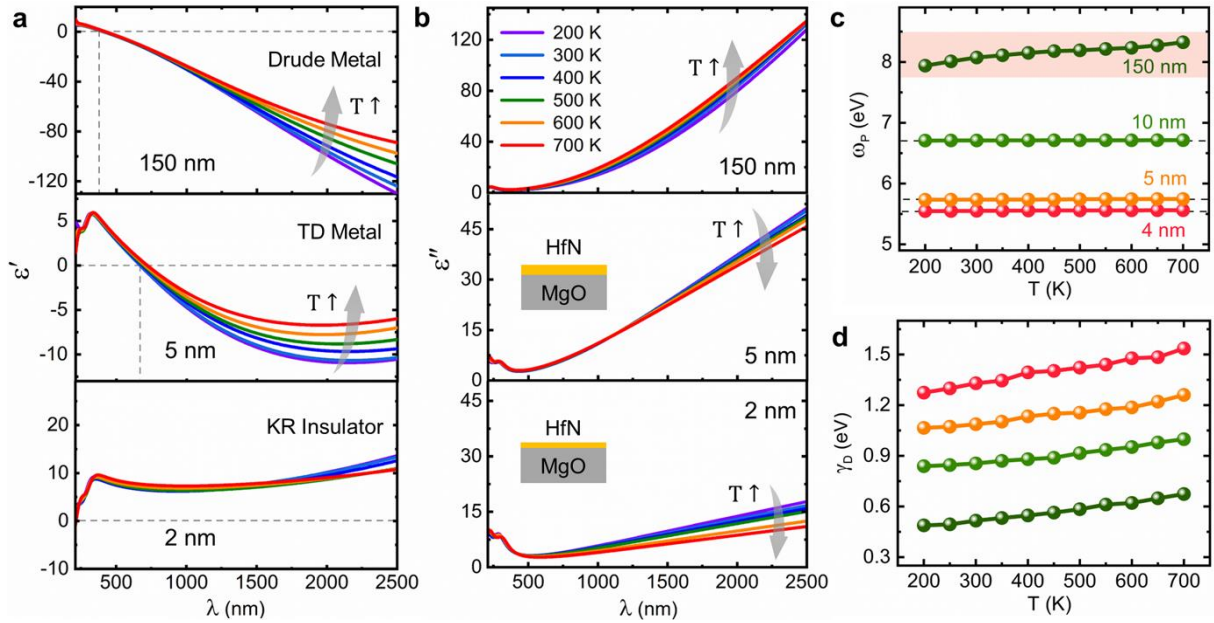

**Fig. 2 | Temperature-dependent dielectric permittivity of HfN across three different optical regimes.** (**a**) Real ($\varepsilon'$) and (**b**) imaginary ($\varepsilon''$) components of dielectric permittivity of HfN film as functions of temperature for three different thickness values, 150 nm, 5 nm, and 2 nm, representing the Drude metallic, TD-plasmonic and KR-insulator regimes, respectively. Temperature-dependent (**c**) plasma frequency and (**d**) Drude damping coefficient of HfN in the TD-plasmonic (with film thickness of 4 nm, 5 nm, 10 nm) and Drude-metallic (150 nm) regime. For TD films, plasma frequency remains constant over a broad temperature range. However, the Drude damping coefficient increases with increase in temperature for all the films.

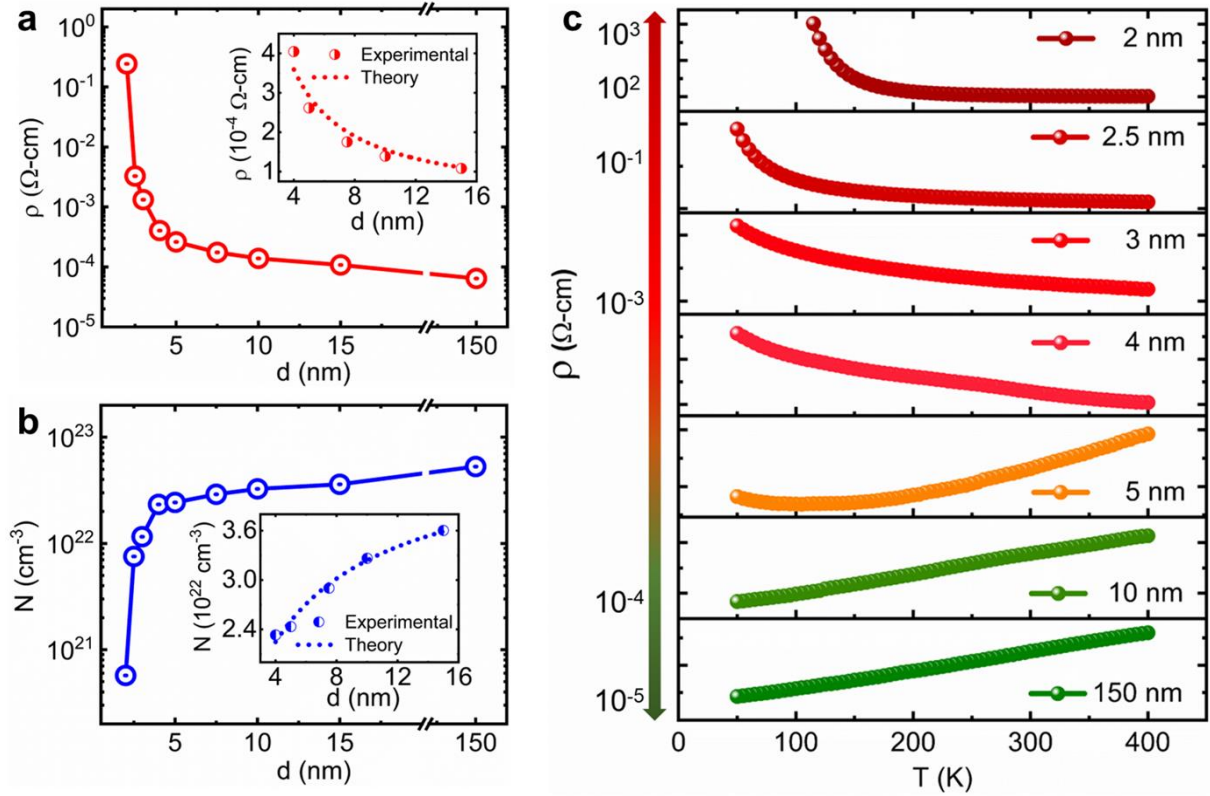

**Fig. 3 | Thickness-dependent electronic metal-insulator transitions in HfN films**. Room temperature (**a**) electrical resistivity and (**b**) carrier concentration as a function of HfN film thickness. With decreasing thickness, resistivity increases and carrier concentration decreases. Insets show resistivity and carrier concentration in the TD-plasmonic regime that are fitted with theoretical equation S22 and S23 (see SI). (**c**) Temperature-dependent electrical resistivity of films with different thicknesses. For 5 nm HfN film, a clear temperature-driven metal-insulator transition is observed.

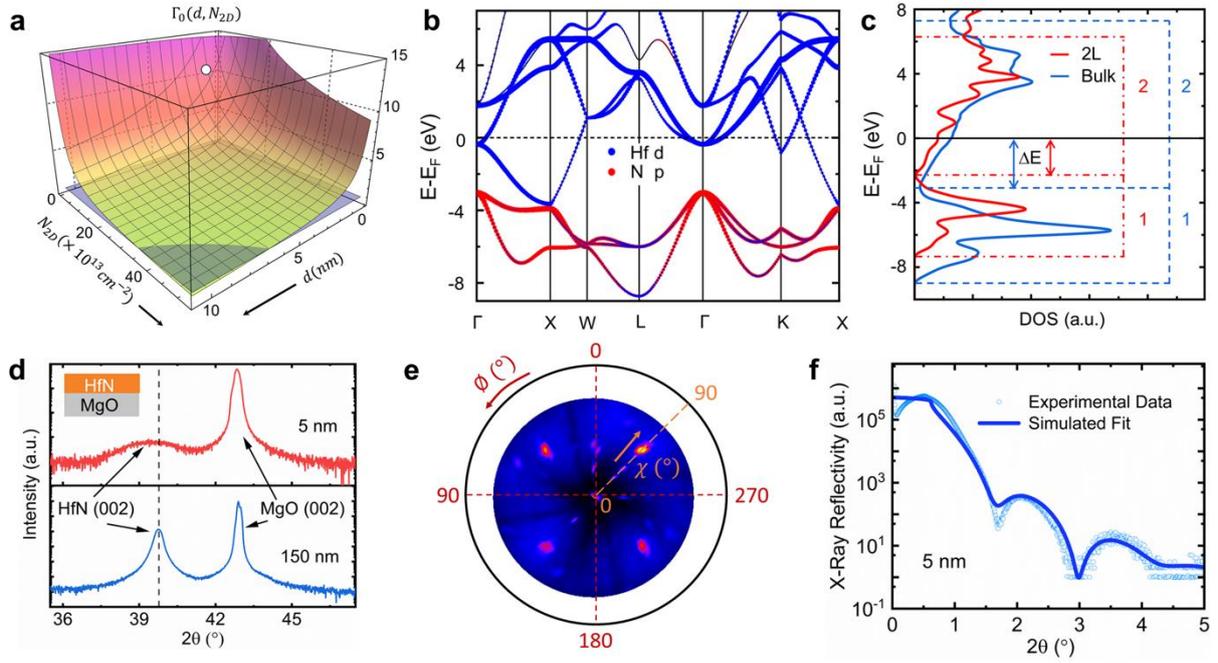

**Fig. 4 | Theoretical calculations and structural characterization of HfN. (a)** $\Gamma_0(n, d)$ calculated with Keldysh-Rytova (KR) interacting potential and with experimental parameters at room temperature shows greater than 1 value for 2-nm-thick HfN (a dot in the image). **(b)** Projected electronic band structure of HfN with $d$-orbitals of Hf (blue) and $p$-orbitals of N (red). **(c)** Electronic density of states for two-layer (2L) (red) and bulk (sky blue) HfN. Dashed lines represent the energy position of the two bands (labelled by 1 and 2) with respect to the Fermi level ($E_F$). **(d)** High-resolution X-ray diffractogram (HRXRD) of 5 and 150 nm HfN films deposited on (001) MgO substrates. **(e)** Pole figure corresponding to the (111) reflection and **(f)** X-ray reflectivity (XRR) spectra of the epitaxial 5 nm film along with the fitting are presented that highlight uniform and coherent epitaxial growth.

**Main References**

1. Brongersma, M. L. & Shalaev, V. M. The Case for Plasmonics. *Science.* **328**, 440–441 (2010).

2. Polman, A. & Atwater, H. A. Plasmonics: optics at the nanoscale. *Mater. Today* **8**, 56 (2005).

3. Zheludev, N. I. & Kivshar, Y. S. From metamaterials to metadevices. *Nat. Mater.* **11**, 917–924 (2012).

4. Ebbesen, T. W., Genet, C. & Bozhevolnyi, S. I. Surface-plasmon circuitry. *Phys. Today* **61**, 44–50 (2008).

5. Fox, M. (2001) Optical Properties of Solids. Oxford University Press, New York.

6. Das, P., Maurya, K. C., Schroeder, J. L., Garbrecht, M. & Saha, B. Near-UV-to-Near-IR Hyperbolic Photonic Dispersion in Epitaxial (Hf,Zr)N/ScN Metal/Dielectric Superlattices. *ACS Appl. Energy Mater.* **5**, 3898–3904 (2022).

7. Das, P., Biswas, B., Maurya, K. C., Garbrecht, M. & Saha, B. Refractory Plasmonic Hafnium Nitride and Zirconium Nitride Thin Films as Alternatives to Silver for Solar Mirror Applications. *ACS Appl. Mater. Interfaces* **14**, 46708–46715 (2022).

8. Bondarev, I. V. & Shalaev, V. M. Universal features of the optical properties of ultrathin plasmonic films. *Opt. Mater. Express* **7**, 3731 (2017).

9. L. V. Keldysh. Coulomb interaction in thin semiconductor and semimetal films. *Pis'ma Zh. Eksp. Teor. Fiz.* **29**, 716–719 (1979) [Engl. translation: *JETP Lett.* **29**, 658–661 (1980)].

10. Boltasseva, A. & Shalaev, V. M. Transdimensional Photonics. *ACS Photonics* **6**, 1–3

(2019).

11. Halas, N. J., Lal, S., Chang, W.-S., Link, S. & Nordlander, P. Plasmons in Strongly Coupled Metallic Nanostructures. *Chem. Rev.* **111**, 3913–3961 (2011).

12. Zhang, X., Chen, Y. L., Liu, R.-S. & Tsai, D. P. Plasmonic photocatalysis. *Reports Prog. Phys.* **76**, 046401 (2013).

13. Guler, U. *et al.* Local Heating with Lithographically Fabricated Plasmonic Titanium Nitride Nanoparticles. *Nano Lett.* **13**, 6078–6083 (2013).

14. Atwater, H. A. & Polman, A. Plasmonics for improved photovoltaic devices. *Nat. Mater.* **9**, 205–213 (2010).

15. Homola, J., Yee, S. S. & Gauglitz, G. Surface plasmon resonance sensors: review. *Sensors Actuators B Chem.* **54**, 3–15 (1999).

16. García de Abajo, F. J. Optical excitations in electron microscopy. *Rev. Mod. Phys.* **82**, 209–275 (2010).

17. Leuthold, J. *et al.* Plasmonic Communications: Light on a Wire. *Opt. Photonics News* **24**, 28 (2013).

18. Kauranen, M. & Zayats, A. V. Nonlinear plasmonics. *Nat. Photonics* **6**, 737–748 (2012).

19. Bigot, J.-Y., Halté, V., Merle, J.-C. & Daunois, A. Electron dynamics in metallic nanoparticles. *Chem. Phys.* **251**, 181–203 (2000).

20. Li, H. Y. *et al.* Analysis of the Drude model in metallic films. *Appl. Opt.* **40**, 6307 (2001).

21. Maurya, K. C. *et al.* Polar Semiconducting Scandium Nitride as an Infrared Plasmon and Phonon–Polaritonic Material. *Nano Lett.* **22**, 5182–5190 (2022).

22. Grigorenko, A. N., Polini, M. & Novoselov, K. S. Graphene plasmonics. *Nat. Photonics*

**6**, 749–758 (2012).

23. Ouyang, Q. *et al.* Sensitivity Enhancement of Transition Metal Dichalcogenides/Silicon Nanostructure-based Surface Plasmon Resonance Biosensor. *Sci. Rep.* **6**, 28190 (2016).

24. Li, X., Tung, C. H. & Pey, K. L. The nature of dielectric breakdown. *Appl. Phys. Lett.* **93**, 1–4 (2008).

25. Abd El-Fattah, Z. M. *et al.* Plasmonics in Atomically Thin Crystalline Silver Films. *ACS Nano* **13**, 7771–7779 (2019).

26. Kossoy, A. *et al.* Optical and Structural Properties of Ultra-thin Gold Films. *Adv. Opt. Mater.* **3**, 71–77 (2015).

27. Naik, G. V., Shalaev, V. M. & Boltasseva, A. Alternative Plasmonic Materials: Beyond Gold and Silver. *Adv. Mater.* **25**, 3264–3294 (2013).

28. Zgrabik, C. M. & Hu, E. L. Optimization of sputtered titanium nitride as a tunable metal for plasmonic applications. *Opt. Mater. Express* **5**, 2786 (2015).

29. Shah, D. *et al.* Thickness-Dependent Drude Plasma Frequency in Transdimensional Plasmonic TiN. *Nano Lett.* **22**, 4622–4629 (2022).

30. Shah, D., Reddy, H., Kinsey, N., Shalaev, V. M. & Boltasseva, A. Optical Properties of Plasmonic Ultrathin TiN Films. *Adv. Opt. Mater.* **5**, 1–5 (2017).

31. Mei, A. B., Rockett, A., Hultman, L., Petrov, I. & Greene, J. E. Electron/phonon coupling in group-IV transition-metal and rare-earth nitrides. *J. Appl. Phys.* **114**, (2013).

32. Chung, S. *et al.* Nanosecond long excited state lifetimes observed in hafnium nitride. *Sol. Energy Mater. Sol. Cells* **169**, 13–18 (2017).

33. Chakraborty, S. *et al.* Phononic bandgap and phonon anomalies in HfN and HfN/ScN

metal/semiconductor superlattices measured with inelastic x-ray scattering. *Appl. Phys. Lett.* **117**, 1–6 (2020).

34. Bondarev, I. V., Mousavi, H. & Shalaev, V. M. Optical response of finite-thickness ultrathin plasmonic films. *MRS Commun.* **8**, 1092–1097 (2018).

35. Reddy, H. *et al.* Temperature-Dependent Optical Properties of Plasmonic Titanium Nitride Thin Films. *ACS Photonics* **4**, 1413–1420 (2017).

36. Vertchenko, L. *et al.* Cryogenic characterization of titanium nitride thin films. *Opt. Mater. Express* **9**, 2117 (2019).

37. Van Bui, H., Kovalgin, A. Y. & Wolters, R. A. M. On the difference between optically and electrically determined resistivity of ultra-thin titanium nitride films. *Appl. Surf. Sci.* **269**, 45–49 (2013).

38. Taylor, K. M. & Lenie, C. Some Properties of Aluminum Nitride. *J. Electrochem. Soc.* **107**, 308 (1960).

39. Bartzsch, H., Glöß, D., Böcher, B., Frach, P. & Goedicke, K. Properties of SiO2 and Al2O3 films for electrical insulation applications deposited by reactive pulse magnetron sputtering. *Surf. Coatings Technol.* **174**–**175**, 774–778 (2003).

40. Roy, M., Mucha, N. R., Fialkova, S. & Kumar, D. Effect of thickness on metal-to-semiconductor transition in 2-dimensional TiN thin films. *AIP Adv.* **11**, (2021).

41. Platzman, P. M. & Fukuyama, H. Phase diagram of the two-dimensional electron liquid. *Phys. Rev. B* **10**, 3150–3158 (1974).

42. Mukhopadhyay, D., Rudra, S., Biswas, B., Das, P. & Saha, B. Surface scattering-dependent electronic transport in ultrathin scandium nitride films. *Appl. Phys. Lett.* **123**, (2023).

43. Rudra, S., Rao, D., Poncé, S. & Saha, B. Reversal of Band-Ordering Leads to High Hole Mobility in Strained p-type Scandium Nitride. *Nano Lett.* **23**, 8211–8217 (2023).

44. Catellani, A. & Calzolari, A. Plasmonic properties of refractory titanium nitride. *Phys. Rev. B* **95**, 115145 (2017).

45. Biswas, B. *et al.* Magnetic Stress-Driven Metal-Insulator Transition in Strongly Correlated Antiferromagnetic CrN. *Phys. Rev. Lett.* **131**, 126302 (2023).

46. Mori-Sánchez, P., Cohen, A. J. & Yang, W. Localization and Delocalization Errors in Density Functional Theory and Implications for Band-Gap Prediction. *Phys. Rev. Lett.* **100**, 146401 (2008).

47. Jiang, H. First-principles approaches for strongly correlated materials: A theoretical chemistry perspective. *Int. J. Quantum Chem.* **115**, 722–730 (2015).

48. Wager, J. F. Real- and reciprocal-space attributes of band tail states. *AIP Adv.* **7**, (2017).

49. Timrov, I., Vast, N., Gebauer, R. & Baroni, S. turboEELS—A code for the simulation of the electron energy loss and inelastic X-ray scattering spectra using the Liouville–Lanczos approach to time-dependent density-functional perturbation theory. *Comput. Phys. Commun.* **196**, 460–469 (2015).

50. Saha, B., Acharya, J., Sands, T. D. & Waghmare, U. V. Electronic structure, phonons, and thermal properties of ScN, ZrN, and HfN: A first-principles study. *J. Appl. Phys.* **107**, 033715 (2010).

# Methods

**Growth Details**

HfN films are deposited on (001) MgO substrate (1 cm × 1 cm) inside an ultrahigh vacuum chamber with a base pressure of $1 \times 10^{-9}$ Torr using a reactive DC magnetron sputtering system (PVD Products, Inc.). The Hf target has the dimensions of 2 inch in. diameter and 0.25 inch in. thickness. Before the deposition, substrates are cleaned with acetone and methanol for 15 minutes. HfN films with different thicknesses are deposited at a pressure of 5 mTorr maintaining an Ar: $N_2$ gas mixture ratio of 9:2 sccm. During the deposition, the target power and the substrate temperature are maintained at a constant 100 W and 900℃, respectively. For the transmission measurements, HfN films are also deposited on double-sided polished quartz substrate along with the (001) MgO substrates.

**Characterization Methods**

The optical properties of HfN films are measured in reflection mode at three different incident angles (55º, 65º, 75º) using a variable angle spectroscopic ellipsometer (VASE) (J.A. Woollam Co.). The experimental Psi ($\psi$) and Delta ($\Delta$) spectra were fitted using a Drude-Lorentz oscillator model in CompleteEASE software. Temperature-dependent optical properties are measured from 200K to 700K at 70° angle of incidence using a vacuum-based cryostat attached with the ellipsometer. Angle-dependent reflection and transmission measurements are also performed with the VASE machine. Dielectric permittivities are extracted from the fitting of experimental Psi ($\psi$) and delta ($\Delta$) spectra with the Drude and Lorentz oscillator model. The Drude term considers the contribution from conduction electrons and the Lorentz term corresponds to the interband transitions. The total permittivity can be expressed as,

$$\varepsilon = \varepsilon' + i\varepsilon'' = \varepsilon_{core} - \frac{\omega_p^2}{\omega^2 + i\gamma_D \omega} + \sum_{j=1}^{2} \frac{\omega_{l,j}^2}{\omega_{0,j}^2 - \omega^2 - i\gamma_j \omega}$$

Where, $\varepsilon_{core}$, $\omega_p$, and $\gamma_D$ are core dielectric constant, unscreened plasma frequency, and the Drude damping coefficient, respectively. $\omega_{l,j}^2 = f_j \omega_{0,j} \gamma_j$ where, $f_j$, $\omega_{0,j}$, and $\gamma_j$ describe the Lorentz oscillator strength, resonant energy, and the damping coefficient, respectively. The unscreened plasma frequency is given by the formula, $\omega_p = \sqrt{\frac{4\pi N e^2}{m^* \varepsilon_0}}$, whereas the screened plasma frequency depends on the core dielectric constant ($\varepsilon^{core}$) of the medium, $\omega_p = \sqrt{\frac{4\pi N e^2}{m^* \varepsilon^{core}}}$. Here, $N$, $m^*$, and $\varepsilon_0$ are the electron concentration, effective mass, and free space permittivity, respectively.

Normal transmission measurements are performed from 200 nm to 2500 nm spectral range using a Cary 5000 Agilent UV-Vis-NIR spectrophotometer. The substrate (quartz) is used as a baseline for the measurement.

Room temperature electrical resistivity and electron concentrations of the HfN films are measured using the Ecopia HMS-3000 Hall measurement system. Temperature-dependent electrical resistivity measurements are performed from 50K to 400K in standard four-probe geometry using a cryogen-free quantum device physical property measurement system (PPMS).

High-resolution X-ray diffraction (HRXRD) is performed with Rigaku smartlab X-ray diffractometer. The rotating anode X-ray generator is set at 4.5 kW during the measurement. Parallel beam optics with a multi-layer X-ray mirror, a Germanium (220) 2-bounce channel cut monochromator, and a Germanium (220) 2-bounce analyser are utilized for the measurement. XRR fitting was performed with smartlab software using Parratt formalism.

**Acknowledgements**

P.D. and B.S. acknowledge the International Centre for Materials Science (ICMS) and Sheikh Saqr Laboratory (SSL) of Jawaharlal Nehru Centre for Advanced Scientific Research (JNCASR) for support. B.S. acknowledges Ras Al Khaimah Centre for Advanced Materials and Raknor LLC for financial support. B.S. acknowledges Young Scientist Research Award (YSRA) from the Board of Research in Nuclear Sciences (BRNS), Department of Atomic Energy (DAE), India with grant number 59/20/10/2020-BRNS/59020 for partial financial support. *P.D., D.R., and B.S. acknowledge the* SAMat *Research Facilities, JNCASR, Bengaluru.* P.D., S.R., and S.B. thanks CSIR for the fellowship. D.R. thanks JNCASR for the fellowship. S.R. and B.S. acknowledged the PARAM Yukti supercomputing facility under the national supercomputing mission of India at JNCASR, Bengaluru for providing computational resources. M.G. and A.I.K.P. acknowledge the facilities of Sydney Microscopy and Microanalysis at the University of Sydney. I.V.B. gratefully acknowledges support from the U.S. Army Research Office under award No. W911NF2310206. A.B. and V.M.S. acknowledge support by the U.S. Department of Energy, Office of Basic Energy Sciences, Division of Materials Sciences and Engineering under Award DE-SC0017717 and Air Force Office of Scientific Research (AFOSR) under award FA9550-20-1-0124. P.D. thanks Dr. Bidesh Biswas for various discussions.

**Author contributions:** P.D. and B.S. conceived this project. P.D. deposited the thin films and performed optical characterizations. S.R. and B.S. performed the theoretical modelling and analysis. D.R. and S.B. performed the electrical measurements. S.B. performed magnetoresistance measurements. A.I.K.P. performed the TEM sample preparation and M.G. performed the TEM imaging and analysis. I.V.B. developed the 2D theoretical model for the case of ultrathin finite-thickness TD films. A.B., I.V.B., and V.M.S. provided valuable insight into interpretations of the experimental results. All authors discussed and contributed to the preparation of the manuscript.

**Competing Interests:** There are no competing interests to declare.

**Data and materials availability:** All data required to evaluate the conclusions in the manuscript are available in the main text or the supplementary information.

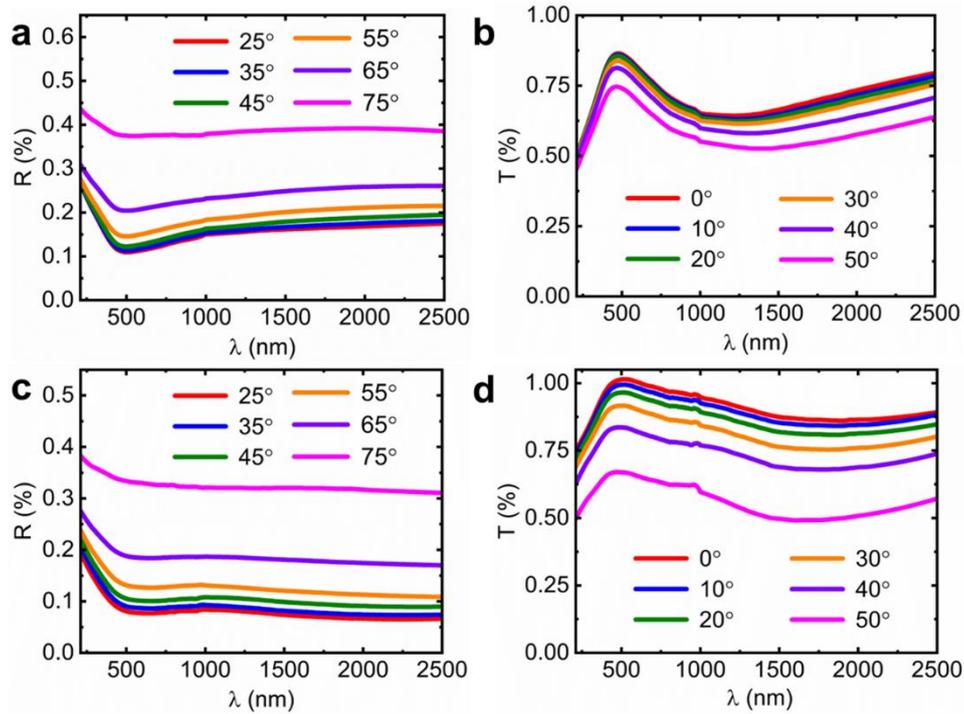

**Extended Data Fig. 1 | Angle-dependent reflection and transmission measurements**. Angle-dependent (**a, c**) reflection and (**b, d**) transmission spectra of 5 and 2 nm HfN film, respectively. Reflection measurements are performed from 25° to 75° in the 210-2500 nm wavelength range using spectroscopic ellipsometry, while transmission measurements are performed from 0° to 50°. The reflection of 5 nm and 2 nm films increases with an increase in the angle of incidence, while transmission decreases with the angle of incidence. However, the dip in the reflection spectra and the peak in the transmission spectra do not change.

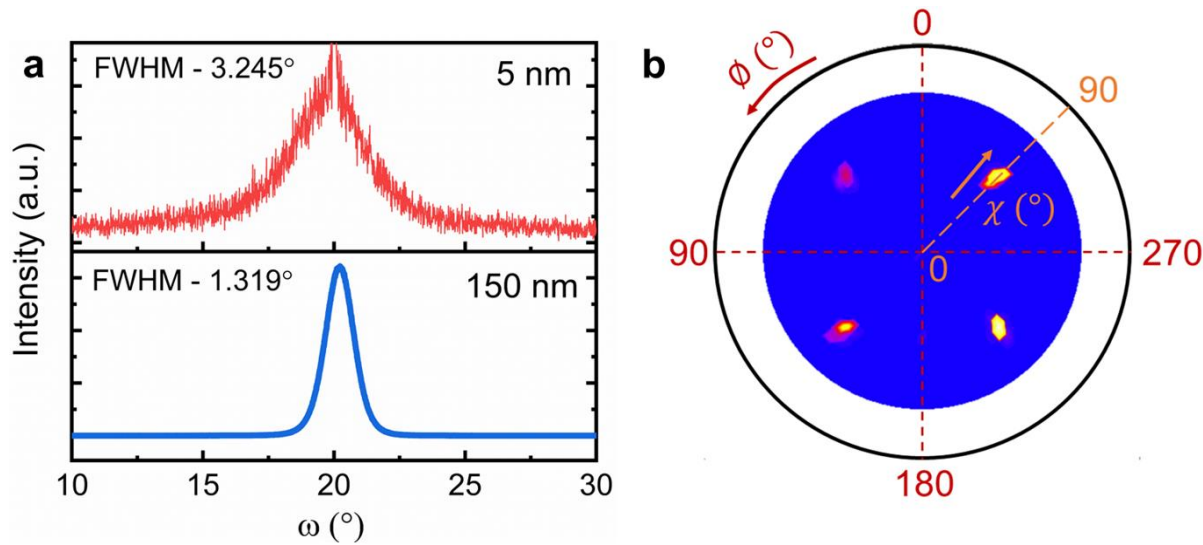

**Extended Data Fig. 2 | X-rad diffraction of bulk HfN film.** (**a**) Rocking curve (ω-scan) of HfN (002) plane for 5 and 150 nm thickness. The full-width-half-maximum (FWHM) of the rocking curve serves as a quantifying parameter for the crystalline growth. The FWHM of the bulk HfN film (thickness 150 nm) is 1.32°, which is significantly low demonstrating the highly crystalline growth of the bulk film. However, the FWHM increases in the ultrathin film showing the trend of decreasing crystallinity with thickness. (**b**) Pole figure corresponds to (111) reflection of 150 nm HfN film. Four distinguished spots which are 90° apart signifies the epitaxial growth on the (001) MgO substrate.

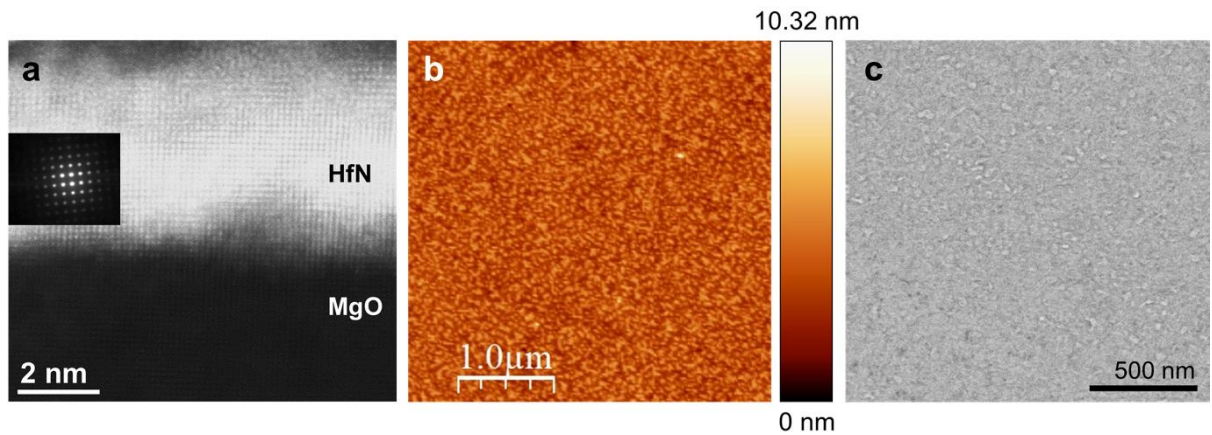

**Extended Data Fig. 3 | Microscopy images of HfN films.** (**a**) High-resolution STEM image of the HfN/MgO interface exhibiting cubic epitaxial growth of the 5 nm film on the MgO substrate. The inset shows the electron diffraction pattern of the HfN film. Edge dislocations at the substrate-film interface are visible. (**b**) Atomic Force Microscope image of a 5 nm film demonstrating uniform coverage over the substrate with Root-Mean-Squared (RMS) roughness of 0.98 nm. (**c**) Plan-view FESEM images of the 5 nm HfN films indicate continuous growth of HfN films with some surface features.

**Extended Data Table I: Comparison of the unscreened bulk plasma frequency ($\omega_P$) and Drude damping coefficient ($\gamma_D$) for various plasmonic metals** [7,29,51,52].

| Plasmonic metals | Plasma frequency ($\omega_P$) (eV) | Drude damping coefficient ($\gamma_D$) (eV) |
|---|---|---|
| Ag | 9.30 | 0.07 |
| Au | 9.16 | 0.055 |
| TiN | 7.60 | 0.20 |
| ZrN | 8.26 | 0.275 |
| HfN | 7.85 | 0.52 |

**Extended Data Table II: Room temperature carrier concentration and electrical resistivity of various HfN films.**

| Thickness (nm) | Carrier Concentration (cm$^{-3}$) | Resistivity (Ohm-cm) |
|---|---|---|
| 2 | $3.70 \times 10^{20}$ | $2.41 \times 10^{-1}$ |
| 2.5 | $7.54 \times 10^{21}$ | $3.28 \times 10^{-3}$ |
| 3 | $1.16 \times 10^{22}$ | $1.32 \times 10^{-3}$ |
| 4 | $2.37 \times 10^{22}$ | $4.04 \times 10^{-4}$ |
| 5 | $2.43 \times 10^{22}$ | $2.62 \times 10^{-4}$ |
| 7.5 | $2.52 \times 10^{22}$ | $1.71 \times 10^{-4}$ |
| 10 | $3.26 \times 10^{22}$ | $1.39 \times 10^{-4}$ |
| 15 | $3.60 \times 10^{22}$ | $1.08 \times 10^{-4}$ |
| 150 | $5.30 \times 10^{22}$ | $6.43 \times 10^{-5}$ |

**Extended Data References**

51. Reddy, H. *et al.* Temperature-Dependent Optical Properties of Single Crystalline and Polycrystalline Silver Thin Films. *ACS Photonics* **4**, 1083–1091 (2017).

52. Reddy, H., Guler, U., Kildishev, A. V., Boltasseva, A. & Shalaev, V. M. Temperature-dependent optical properties of gold thin films. *Opt. Mater. Express* **6**, 2776 (2016).

# Supplementary Information

## Electron Confinement-induced Plasmonic Breakdown in Metals

Prasanna Das[1], Sourav Rudra[1], Dheemahi Rao[1], Souvik Banerjee[1], Ashalatha Indiradevi Kamalasanan Pillai[2], Magnus Garbrecht[2], Alexandra Boltasseva[3], Igor V. Bondarev[4], Vladimir M. Shalaev[3], and Bivas Saha[1,5]*

[1]*Chemistry and Physics of Materials Unit and International Centre for Materials Science*, *Jawaharlal Nehru Centre for Advanced Scientific Research, Bangalore 560064, India.*

[2]*Sydney Microscopy and Microanalysis, The University of Sydney, Camperdown, NSW 2006, Australia.*

[3]*Elmore Family School of Electrical and Computer Engineering, and Birck Nanotechnology Centre, Purdue University, West Lafayette, IN 47907, USA*

[4]*Department of Mathematics & Physics, North Carolina Central University, Durham, NC 27707, USA.*

[5]*School of Advanced Materials and Sheikh Saqr Laboratory, Jawaharlal Nehru Centre for Advanced Scientific Research, Bangalore 560064, India.*

*Correspondence to: bsaha@jncasr.ac.in and bivas.mat@gmail.com

# Supplementary Text

**Section I. Surface plasmon polariton parameters of bulk HfN**

Angle-dependent reflection of thick HfN film is measured with spectroscopic ellipsometry from 210 nm to 2500 nm spectral ranges. Reflection spectra show a dip around 415 nm corresponding to its epsilon-near-zero wavelength. In long wavelength regime, reflection is very high and increases with increasing angle of incidence (Fig. S1a). Different surface plasmon polariton (SPP) resonance parameters are computed for HfN from the experimental dielectric permittivity, and compared with TiN and ZrN [7,53]. SPPs propagating at the interface follow the dispersion relation,

$$k_{SPP} = \frac{2\pi}{\lambda}\sqrt{\frac{\varepsilon_m \varepsilon_a}{\varepsilon_m + \varepsilon_a}} \tag{S1}$$

where, $k_{SPP}$, $\varepsilon_m$ and $\varepsilon_a$ denote the propagation constant in the direction of propagation, and the dielectric permittivity of metal and air, respectively. The distance at which the energy of surface plasmon decays by a factor of *1/e* in the direction of propagation is defined as the propagation length, denoted by

$$L = \frac{1}{2\,Im(k_{SPP})} \tag{S2}$$

SPP propagation starts when HfN becomes plasmonic, i.e., after the ENZ wavelength, and the propagation length increases as wavelength increases, exhibiting a large value of 96 µm at 2500 nm (see Fig. S1b). The electromagnetic field of the surface plasmon shows the maximum value at the metal-air interface and decays perpendicularly into the two media. The decay length is defined as the distance from the interface at which the field falls off by *1/e*,

$$\delta = \frac{\lambda}{2\pi} \sqrt{\frac{|\varepsilon'| + \varepsilon_a}{|\varepsilon_m|^2}} \tag{S3}$$

Therefore, at the cross-over wavelength when the real part becomes zero, the large decay length corresponds to very low loss. Plasmon decay length exhibits a value of 32 nm at the ENZ wavelength for HfN, which indicates its good plasmonic response (Fig. S1c). The quality factor for surface plasmon polariton defined by, $Q_{SPP} = \frac{\varepsilon'^2}{\varepsilon''}$, serves as a performance figure-of-merit of metal for plasmonic applications. Results show that HfN exhibits a higher quality factor ($Q_{HfN}$ ~ 92 at $\lambda$= 1500 nm) in the visible-to-near-infrared spectral ranges (Fig. S1d).

**Section II. Transdimensional plasmonics**

**A. Keldysh-Rytova potential**

In a film with thickness $d$ and background dielectric constant $\varepsilon^{core}$, the Coulomb interaction between two point charges $e$ and $e´$ transforms into a pure 2D potential known as the Keldysh-Rytova potential. When the film dielectric constant becomes larger compared to those of surrounding media ($\varepsilon_{1,2}^{core}$) and in-plane inter-charge distance ($\rho$) becomes higher than the film thickness. The potential takes the form,

$$V(\rho) = \frac{\pi e e'}{\varepsilon^{core} d}\left[H_0\left(\frac{\varepsilon_1^{core}+\varepsilon_2^{core}}{\varepsilon^{core}}\frac{\rho}{d}\right) - Y_0\left(\frac{\varepsilon_1^{core}+\varepsilon_2^{core}}{\varepsilon^{core}}\frac{\rho}{d}\right)\right] \quad (S4)$$

Where, $H_0$ and $Y_0$ are Struve and Neumann functions, respectively.

In the interval, $1 \ll \frac{\rho}{d} \ll \frac{\varepsilon^{core}}{\varepsilon_1+\varepsilon_2}$, Keldysh potential is reduced to,

$$V(\rho) = \frac{2ee'}{\varepsilon^{core} d}\left[\ln\left(\frac{2\varepsilon^{core}}{\varepsilon_1^{core}+\varepsilon_2^{core}}\frac{d}{\rho}\right) - C\right] \quad (S5)$$

(C ≈ 0.577 is the Euler constant)

This potential varies much faster than any Coulomb-type potential ($\sim \frac{1}{\rho}$) does.

If the thickness of the film reduces to a value such that the condition $1 \ll \frac{\varepsilon^{core}}{\varepsilon_1^{core}+\varepsilon_2^{core}} \ll \frac{\rho}{d}$ is satisfied then, the Keldysh potential takes the form,

$$V(\rho) = \frac{2ee'}{(\varepsilon_1^{core}+\varepsilon_2^{core})\rho}$$

which does not include any term representing the film material, and therefore, shows no screening leading ultimately to an insulating state [8].

## B. Plasma frequency

Vertical confinement in the transdimensional plasmonic film leads to a spatially dispersive nature of the plasma frequency. When the free electron wave vectors ($k$) are much less than the cut-off wave vector ($k_c$), plasma frequency takes the form,

$$\omega_p = \omega_p(k) = \frac{\omega_p^{3D}}{\sqrt{1+\frac{\varepsilon_1^{core}+\varepsilon_2^{core}}{\varepsilon^{core}kd}}} \tag{S6}$$

Here, $\omega_p^{3D}$ is the bulk plasma frequency, defined as $\omega_p^{3D} = \sqrt{\frac{4\pi Ne^2}{m^*\varepsilon^{core}}}$, where $N=N_{3D}$, $e$, and $m^*$ are the electron concentration, electron charge, and effective mass, respectively.

For relatively thick film $\left(\frac{\varepsilon_1^{core}+\varepsilon_2^{core}}{\varepsilon^{core}kd} \ll 1\right)$, $\omega_p = \omega_p^{3D}$ and for ultrathin film $\left(\frac{\varepsilon_1^{core}+\varepsilon_2^{core}}{\varepsilon^{core}kd} \gg 1\right)$,

$$\omega_p = \omega_p^{2D}(k) = \sqrt{\frac{4\pi e^2 N_{2D}k}{(\varepsilon_1^{core}+\varepsilon_2^{core})m^*}}$$

where, $N_{2D}(=N_{3D}d)$ is the surface electron density.

The spatial dispersion and associated optical properties of thin plasmonic films can be adjusted by varying the film thickness ($d$), material composition ($\varepsilon^{core}$), and choice of the substrates ($\varepsilon_1^{core}$) and coating layers ($\varepsilon_2^{core}$).

## C. Dependence of plasma frequency on temperature and thickness

Considering a plasmonic film is at a thermal equilibrium (T), then the thermal averaging of equation (S6) can be expressed as,

$$\overline{\omega_p}(d, T) = \frac{\int_0^{k_c} dk k \omega_p(k) n(k)}{\int_0^{k_c} dk k n(k)} \tag{S7}$$

where, $n(k) = \frac{1}{e^{\frac{\hbar \omega_p(k)}{k_B T}} - 1}$ is the plasmon mean occupation number (plasmons are bosons).

Here, the numerator sums up all the boson plasma frequency modes with different $k$ and the denominator provides the total number of such modes in the 2D $k$-space (bounded above by the plasmon cut-off parameter $k_c$).

By putting equation (S6) in equation (S7), one obtains

$$\frac{\overline{\omega_p}(T)}{\omega_p^{3D}} = \frac{\int_0^1 dt t \sqrt{\frac{t}{t+\frac{1}{a}}} \Big/ \left[\exp\left(\alpha \sqrt{\frac{t}{t+\frac{1}{a}}}\right) - 1\right]}{\int_0^1 dt t \Big/ \left[\exp\left(\alpha \sqrt{\frac{t}{t+\frac{1}{a}}}\right) - 1\right]} \tag{S8}$$

Where, $t = \frac{k}{k_c}$, $\alpha = \frac{\hbar \omega_p^{3D}}{k_B T}$, and $a = \frac{\varepsilon^{core} k_c d}{\varepsilon_1^{core} + \varepsilon_2^{core}}$

For ultrathin film ($\frac{\varepsilon^{core} k_c d}{\varepsilon_1^{core} + \varepsilon_2^{core}} \ll 1$), the value of $a$ is less than unity and for temperature higher than the cryogenic temperature, $\alpha$ is also less than unity. Therefore, $\alpha \sqrt{\frac{t}{t+\frac{1}{a}}} \sim \alpha \sqrt{at} < 1$.

Using the Taylor series expansion of the exponential term, the integral turns out to be,

$$\frac{\overline{\omega_p}(T)}{\omega_p^{3D}} = \frac{1}{2 \int_0^1 dt \sqrt{t\left(t+\frac{1}{a}\right)}} \tag{S9}$$

The *T*-dependence of the plasma frequency is cancelled out while *d*-dependence prevails. Thus, for transdimensional plasmonic film, plasma frequency becomes temperature-independent for any temperature greater than cryogenic temperatures ($T \sim 10K$). In the cryogenic regime, plasma frequency suddenly drops with lowering the temperature as recently observed in 100 nm TiN film [36].

Completing the integration in equation (S9) results in

$$\overline{\omega_p}(T) = \frac{2C^2 d^2 \omega_p^{3D}}{(1+2Cd)\sqrt{Cd(1+Cd)} - \sinh^{-1}(\sqrt{Cd})} \quad (S10)$$

Where $C = \frac{a}{d} = \frac{\varepsilon^{core} k_c}{\varepsilon_1^{core} + \varepsilon_2^{core}}$, which gives the $\sqrt{d}$ thickness dependence behaviour of the ultrathin plasmonic film [54].

### D. Dielectric permittivity

Spatially dispersive plasma frequency in ultrathin plasmonic film results in non-local Drude dielectric response function,

$$\varepsilon(\omega, k) = \varepsilon^{core}\left(1 - \frac{\omega_p^2(k)}{\omega(\omega + i\gamma_D)}\right) \quad (S11)$$

where, $\omega_p(k)$ and $\gamma_D$ represent the plasma frequency and Drude damping coefficient, respectively.

By substituting equation (S6) in equation (S11), we acquire the thickness-dependent dielectric permittivity-

$$\varepsilon(\omega, k) = \varepsilon^{core}\left(1 - \frac{(\omega_p^{3D})^2}{\omega(\omega + i\gamma_D)\left(1 + \frac{\varepsilon_1^{core} + \varepsilon_2^{core}}{\varepsilon^{core} kd}\right)}\right) \quad (S12)$$

At frequencies well below the interband transition frequencies such that $\omega > \omega_p^{3D}\sqrt{2\pi\varepsilon^{core}d/((\varepsilon_1^{core}+\varepsilon_2^{core})L)}$ [34], the dielectric permittivity can be represented as a non-dispersive (k-independent) isotropic term containing film thickness $d$ and lateral size $L$,

$$\varepsilon(\omega) = \varepsilon^{core}\left(1 - \frac{(\omega_p^{3D})^2}{\omega(\omega+i\gamma_D)\left(1+\frac{L(\varepsilon_1^{core}+\varepsilon_2^{core})}{2\pi\varepsilon d}\right)}\right) \quad (S13)$$

Substituting $x = \frac{\omega}{\omega_p^{3D}}$ and $\varepsilon_{eff} = \frac{\varepsilon_1^{core}+\varepsilon_2^{core}}{\varepsilon^{core}}$ in equation (S12), one gets

$$\frac{\varepsilon(\omega)}{\varepsilon^{core}} = 1 - \frac{1}{x\left(x+i\frac{\gamma_D}{\omega_p^{3D}}\right)\left(1+\frac{\varepsilon_{eff}}{kd}\right)} = 1 - \frac{\left(x-i\frac{\gamma_D}{\omega_p^{3D}}\right)}{x\left(x^2+\left(\frac{\gamma_D}{\omega_p^{3D}}\right)^2\right)\left(1+\frac{\varepsilon_{eff}}{kd}\right)}$$

$$= 1 - \frac{1}{\left(x^2+\left(\frac{\gamma_D}{\omega_p^{3D}}\right)^2\right)\left(1+\frac{\varepsilon_{eff}}{kd}\right)} + i\frac{\left(\frac{\gamma_D}{\omega_p^{3D}}\right)}{x\left(x^2+\left(\frac{\gamma_D}{\omega_p^{3D}}\right)^2\right)\left(1+\frac{\varepsilon_{eff}}{kd}\right)}$$

The real part of the permittivity,

$$\frac{\varepsilon'(\omega)}{\varepsilon^{core}} = 1 - \frac{1}{\left(x^2+\left(\frac{\gamma_D}{\omega_p^{3D}}\right)^2\right)\left(1+\frac{\varepsilon_{eff}}{kd}\right)} = 1 - \frac{(\omega_p^{3D})^2}{(\omega^2+\gamma_D^2)\left(1+\frac{\varepsilon_1^{core}+\varepsilon_2^{core}}{\varepsilon^{core}kd}\right)} \quad (S14)$$

The imaginary part of the permittivity,

$$\frac{\varepsilon''(\omega)}{\varepsilon^{core}} = \frac{\left(\frac{\gamma_D}{\omega_p^{3D}}\right)}{x\left(x^2+\left(\frac{\gamma_D}{\omega_p^{3D}}\right)^2\right)\left(1+\frac{\varepsilon_{eff}}{kd}\right)} = \frac{\gamma_D(\omega_p^{3D})^2}{\omega(\omega^2+\gamma_D^2)\left(1+\frac{\varepsilon_1^{core}+\varepsilon_2^{core}}{\varepsilon^{core}kd}\right)} \quad (S15)$$

For epsilon-near-zero (ENZ) frequency ($\widehat{\omega}_p = \omega(\varepsilon'=0)$ and $x = x_p$),

$$1 - \frac{1}{\left(x_p{}^2 + \left(\frac{\gamma_D}{\omega_p^{3D}}\right)^2\right)\left(1 + \frac{\varepsilon_{eff}}{kd}\right)} = 0$$

$$\text{Or, } \left(x_p{}^2 + \left(\frac{\gamma_D}{\omega_p^{3D}}\right)^2\right) = \frac{1}{\left(1 + \frac{\varepsilon_{eff}}{kd}\right)}$$

$$\text{Or, } x_p{}^2 = \frac{1}{\left(1 + \frac{\varepsilon_{eff}}{kd}\right)} - \left(\frac{\gamma_D}{\omega_p^{3D}}\right)^2$$

$$\text{Or, } \left(\frac{\widehat{\omega_p}}{\omega_p^{3D}}\right)^2 = \frac{1}{\left(1 + \frac{\varepsilon_{eff}}{kd}\right)} - \left(\frac{\gamma_D}{\omega_p^{3D}}\right)^2 \tag{S16}$$

The ratio $\frac{\gamma_D}{\omega_p^{3D}} \ll 1$ and the ENZ frequency is controlled by the thickness of the film *d*.

For the thick films with $\frac{\varepsilon_{eff}}{kd} \ll 1$, the ENZ frequency approaches bulk plasma frequency i.e., $\widehat{\omega_p} \approx \omega_p^{3D}$.

For the ultrathin films with $\frac{\varepsilon_{eff}}{kd} \gg 1$, ENZ frequency becomes very small compared to bulk plasma frequency i.e., $\widehat{\omega_p} \ll \omega_p^{3D}$. With decreasing thickness value, ENZ frequency is red-shifted.

It is clear from equation (S15) that at a fixed frequency, the imaginary part of permittivity decreases with decreasing thickness. However, at ENZ frequency, which varies with the thickness according to equation (S16), imaginary component of the permittivity increases with decreasing *d* due to the damping constant increase.

**E. Condition to achieve metal-insulator transition**

Metal-insulator transition in the ultrathin film occurred when the metallic state of the film, represented by the negative values of the real component of the permittivity ($\varepsilon' < 0$), is transformed into an insulting state, represented by the positive values of the real part of permittivity ($\varepsilon' > 0$). As the permittivity changes with frequency and becomes zero at the ENZ frequency, therefore, thickness-induced metal-insulator transition is considered for a frequency ($\omega$) well below the ENZ frequency.

Therefore, the real part of permittivity can be expressed as, see Equcation (S15),

$$\frac{\varepsilon'}{\varepsilon^{core}} = 1 - \frac{(\omega_p^{3D})^2}{(\omega^2+\gamma_D^2)\left(1+\frac{\varepsilon_1^{core}+\varepsilon_2^{core}}{\varepsilon^{core}kd}\right)} \tag{S17}$$

Therefore, as the thickness $d$ decreases, $\frac{\varepsilon'}{\varepsilon^{core}}$ becomes more positive.

The film will undergo a metal-insulator transition when $\varepsilon' > 0$ i.e.,

$$\varepsilon^{core}\left(1 - \frac{(\omega_p^{3D})^2}{(\omega^2+\gamma_D^2)\left(1+\frac{\varepsilon_1^{core}+\varepsilon_2^{core}}{\varepsilon^{core}kd}\right)}\right) > 0 \tag{S18}$$

### F. Optical resistivity

Optical conductivity $\sigma(\omega)$ is related to the dielectric permittivity through Drude model as,

$$\varepsilon(\omega) = \varepsilon^{core} + i4\pi\frac{\sigma(\omega)}{\omega}$$

Or, $\sigma(\omega) = i\frac{\omega}{4\pi}\left(\varepsilon^{core} - \varepsilon(\omega)\right)$ \hfill (S19)

Using equation (S13), we get

$$\sigma(\omega) = i\frac{\omega}{4\pi}\left(\varepsilon^{core} - \varepsilon^{core}\left(1 - \frac{(\omega_p^{3D})^2}{\omega(\omega+i\gamma_D)\left(1+\frac{L(\varepsilon_1^{core}+\varepsilon_2^{core})}{2\pi\varepsilon d}\right)}\right)\right)$$

$$= \frac{i\varepsilon^{core}(\omega_p^{3D})^2/(4\pi)}{(\omega+i\gamma_D)\left(1+\frac{L(\varepsilon_1^{core}+\varepsilon_2^{core})}{2\pi\varepsilon^{core}d}\right)} \tag{S20}$$

Now, $\omega + i\gamma_D = \omega + i\frac{1}{\tau} = i\left(\frac{1-i\omega\tau}{\tau}\right)$ where $\tau$ is the electron relaxation time.

Therefore, $\sigma(\omega) = \frac{\varepsilon^{core}\tau(\omega_p^{3D})^2/(4\pi)}{(1-i\omega\tau)\left(1+\frac{L(\varepsilon_1^{core}+\varepsilon_2^{core})}{2\pi\varepsilon^{core}d}\right)}$ (S21)

DC conductivity limit, $\sigma_{dc} = \sigma(\omega \to 0) = \frac{\varepsilon^{core}\tau(\omega_p^{3D})^2/(4\pi)}{\left(1+\frac{L(\varepsilon_1^{core}+\varepsilon_2^{core})}{2\pi\varepsilon^{core}d}\right)}$

Optical resistivity, $\rho_{opt} = \frac{1}{\sigma_{dc}} = 4\pi\frac{1+\frac{L(\varepsilon_1^{core}+\varepsilon_2^{core})}{2\pi\varepsilon^{core}d}}{\varepsilon^{core}\tau(\omega_p^{3D})^2} = A\left(1+\frac{B}{d}\right)$ (S22)

Where, $A = \frac{4\pi}{\varepsilon^{core}\tau(\omega_p^{3D})^2}$ and $B = \frac{L(\varepsilon_1^{core}+\varepsilon_2^{core})}{2\pi\varepsilon^{core}}$ are constants.

For metallic films, electrical and optical resistivity are almost equal. Therefore, the volumetric electron concentration can be deduced as,

$N_{3D} = \frac{1}{\rho_{opt}\mu e} = \frac{1}{\mu e A\left(1+\frac{B}{d}\right)} = \frac{1}{A'\left(1+\frac{B}{d}\right)}$ (S23)

Where, $A' = \mu e A$ is constant and variation of mobility $\mu$ with thickness is negligible.

### G. Drude damping

Drude damping arises in thin film due to various scattering phenomena e.g., electron-electron scattering ($\gamma_{e-e}$), electron-phonon scattering ($\gamma_{e-ph}$), surface scattering ($\gamma_{surf}$), grain-boundary scattering ($\gamma_{gb}$), and defect scattering ($\gamma_d$) –

$\gamma_D = \gamma_{e-e} + \gamma_{e-ph} + \gamma_{surf} + \gamma_{gb} + \gamma_d$ (S24)

In single-crystalline HfN films, grain boundary and defect scattering are negligible. In bulk HfN film, electron-electron and electron-phonon scattering dominate, while in ultrathin HfN film, an additional surface scattering is introduced due to the decreased mean free path of electrons in the vertical direction of the film.

$$\gamma_{surf} = \frac{\hbar v_f}{l_{eff}} \tag{S25}$$

Where, $v_f$ is the Fermi velocity, defined as $v_f = \frac{\omega_p}{k_c}$ ($k_c$ is the plasmon cut-off wavevector) and $l_{eff}$ is the effective mean free path defined as,

$$l_{eff} = d\left(ln\left(\frac{l_{bulk}}{d}\right) + 1\right) \tag{S26}$$

($l_{bulk}$ is the electron mean free path in the bulk film).

Surface roughness also contributes to the Drude damping in ultrathin film, where it is expected to vary inversely with the thickness of the film. Therefore, the overall Drude damping coefficient can be written as,

$$\gamma_D = \gamma_{D\,bulk} + \frac{\hbar v_f}{d\left(ln\left(\frac{l_{bulk}}{d}\right)+1\right)} + \frac{C}{d} \tag{S27}$$

Where C is a fitting constant to account for the surface roughness effect and $\gamma_{D\,bulk}$ is the Drude damping value in the bulk film [29].

## H. Temperature dependence of dielectric permittivity

The variation of the real and imaginary parts of dielectric permittivity with temperature for HfN film with three different thicknesses can be understood from the Drude theory. For the bulk HfN film, the unscreened plasma frequency ($\omega_p^{3D}$) and Drude damping coefficient ($\gamma_D$) both increase with the increase in temperature. The bulk plasma frequency depends on the electron concentration ($N_{3D}$) and effective mass ($m^*$) through the relation $\omega_p^{3D} = \sqrt{\frac{4\pi N_{3D} e^2}{m^* \varepsilon_0}}$ [52].

The electron concentration reduces marginally due to volume expansion while electron effective mass decreases with increasing temperature. The increasing trend of plasma frequency with temperature (Fig. 2c) suggests a decrement in effective mass as the dominant mechanism [35]. The Drude damping coefficient relates to the different scattering phenomena which get increased with an increase in temperature.

The dielectric permittivity of the bulk HfN is given by the Drude relation,

$$\varepsilon = \varepsilon_{core} - \frac{\omega_p^2}{\omega^2 + i\gamma_D \omega};$$

Real part, $\varepsilon' = \varepsilon_{core} - \frac{\omega_p^2}{\omega^2 + \gamma_D^2}$ ; and imaginary part, $\varepsilon'' = \frac{\omega_p^2 \gamma_D}{\omega^3 + \gamma_D^2 \omega}$ ;

As the plasma frequency $\omega_p^{3D}$ and Drude damping coefficient $\gamma_D$ increase with temperature, the imaginary part of permittivity of bulk HfN film increases with temperature as shown in Fig. 2c.

Generally, the real part of permittivity is associated with $\omega_p^{3D}$ and is considered to be unaffected by the $\gamma_D$. However, this is obeyed only when $\gamma_D \ll \omega$ (In case of noble metal $\gamma_D \approx 0.02$-$0.05$ eV). When $\gamma_D$ becomes comparable to energies in near-IR spectrum (For HfN, $\gamma_D$ is 0.52 eV and 0.67 eV respectively at 300 $K$ and 700 K), its contribution to the real part of permittivity is prominent [51]. An increase in $\gamma_D$ leads to a decrease in $\varepsilon'$.

For a transdimensional metallic HfN film, the plasma frequency $\omega_p$ becomes independent of temperature as predicted by theory and observed from the experiment [36]. Therefore the variation of $\varepsilon'$ and $\varepsilon''$ with temperature is controlled by the $\gamma_D$. As $\gamma_D$ increases with temperature, magnitude of $\varepsilon'$ decreases. The imaginary part of permittivity $\varepsilon''$ effectively varies with $\gamma_D$ as $\varepsilon'' \sim \frac{1}{\gamma_D}$. Therefore, with increasing temperature, $\varepsilon''$ decreases. Similarly, $\varepsilon''$ of 2 nm HfN film decreases with temperature.

**Section III. Temperature-dependent resistivity fitting of HfN film**

The temperature-dependent transport properties of 5 nm ultrathin HfN films are explained in terms of different conduction mechanisms (Fig. S3). For temperature greater than 250K, resistivity increases proportionally with the temperature indicating its metallic character. For the temperature range 105K < T < 250K, resistivity increases proportionally with the higher order of temperature. Below 105K, the conduction is dominated by two variable range hopping mechanisms. Nearest neighbour hopping (NNH) involves the transition of charge carriers between two nearest neighbours while Mott variable range hopping (MVRH) considers the hopping of electrons between localized states in the vicinity of the Fermi level [40]. For the temperature range of 85-105K, the resistivity can be fitted with the NNH model. The equation for NNH is given by,

$$\rho = \rho_T \, exp(\frac{E_a^{NNH}}{k_B T}) \qquad (S28)$$

Where, $\rho_T$ is a temperature-independent constant and $E_a^{NNH}$ and $k_B$ are activation energy and Boltzmann constant, respectively.

The MVRH model is used to fit the resistivity data in the very low-temperature range of 50-85K. The MVRH equation for 2D semiconductor thin film is given by,

$$\rho = \rho_T \, exp\left(\frac{T_M}{T}\right)^{\frac{1}{3}} \qquad (S29)$$

Where, $T_M$ is the Mott temperature, which relates to the localization strength $L_M$ and Bohr radius $r_{Bohr}$ as

$$T_M = \frac{18}{L_M^3 N(E_F) k_B} \qquad (S30)$$

$$\bar{L}_M = r_{Bohr} \left(\frac{T_M}{T}\right)^{\frac{1}{3}} \qquad (S31)$$

The term $N(E_F)$ is the density of states (DOS) at the Fermi level.

Temperature-dependent resistivity of 2 nm film is fitted with nearest neighbour hopping and Mott variable range hopping mechanism (Fig. S4). For the temperature range 245-400 K, The NNH model is used for the fitting whereas the MVRH model is used for the temperature range 115-245 K.

Temperature-dependent electrical resistivity of ultrathin HfN film of various thicknesses (2nm, 2.5 nm, 3 nm) with and without in-plane magnetic field is shown in a broad temperature range (2 K to 300 K) (Fig. S5). In very low temperatures, ultrathin HfN films show negative magnetoresistance as shown in Table S7. Although the change in magnetoresistance for 2 nm and 2.5 mm is very small and almost unchanged in higher temperature limit, the 3 nm film maintains overall an average of -5.8% negative magnetoresistance in the measured temperature limit suggesting a strong confinement of electrons in 2 nm and 2.5 nm ultrathin HfN films [45,55].

**Section IV. Calculation of Platzman-Fukuyama (PF) ratio**

The mean electron kinetic energy per particle $\langle E_{kin} \rangle$ at zero temperature can be obtained by integrating over the 2D Fermi surface (the circular zone of radius $k_F$, the Fermi momentum absolute value, in the 2D reciprocal space). This gives $\langle E_{kin} \rangle = \hbar^2 \pi N_{2D}/(2m^*)$, where $N_{2D} = N_{3D} d = 1/(\pi \bar{\rho}^2)$ is the surface charge density (its volume counterpart is $N_{3D}$).

The dimensionless Platzman-Fukuyama (PF) ratio of potential interaction energy to mean kinetic energy can be written as, $\Gamma_0(\tilde{d}, n) = \frac{V_{KR}(\bar{\rho})}{\langle E_{kin} \rangle} = \frac{1}{\bar{\varepsilon}\tilde{\varepsilon}\tilde{d}n}\left\{H_0\left(\frac{1}{\tilde{\varepsilon}\tilde{d}\sqrt{\pi n}}\right) - Y_0\left(\frac{1}{\tilde{\varepsilon}\tilde{d}\sqrt{\pi n}}\right)\right\}$ (see the main text for notation used). To compute it, we use $\varepsilon^{core} = 5$ (average of the data presented in Table S1, left column), $\varepsilon_1^{core} = 3$ (MgO dielectric constant) and $\varepsilon_2^{core} = 1$ (air).

Calculated $\Gamma_0(n, d)$ increases drastically as the thickness of the films and the corresponding carrier concentrations decrease rapidly. This suggests a strong electron confinement in the TD plasmonic and KR insulator regimes. Interestingly, for the thinnest 2 nm film, $\Gamma_0(n, d)$ exceeds 1 satisfying the well-established PF condition for Wigner crystallization (WC)

$$\Gamma_0 = \frac{V_{KR}(\bar{\rho})}{\langle E_{kin} \rangle} > 1$$

Here, $V_{KR}(\bar{\rho})$ is the repulsive KR interaction potential introduced in Section 4A, and $\langle E_{kin} \rangle$ is the mean electron kinetic energy per particle.

The PF theory is a strictly 2D microscopical model that relies on an intuitive picture of what will occur when the electron Wigner solid (WS) melts [41]. As solids support transverse phonon modes and liquids do not, one can assume that an instability of the transverse phonon mode could signal the onset of WS melting. Based on this assumption, the PF model parametrizes the vibrational spectrum of the 2D electron WS by a transverse linearly dispersive mode and a longitudinal plasmon mode. It predicts $\Gamma_0 = 2.8$ from the onset of melting indicated by the transverse phonon mode break-up. Because the TD regime contains the film thickness as a parameter of otherwise 2D theoretical description of ultrathin films, a simple rescaling of the coupling constant as follows

$$e^2 \rightarrow \frac{1}{\bar{\varepsilon}(\tilde{\varepsilon}dk_D+1)}e^2 \tag{S32}$$

brings the 2D plasma frequency squared of the PF model to the TD form and thereby generalizes the PF model to include the TD material case. Here, $k_D = k_F\sqrt{2} = \sqrt{4\pi N_{2D}} = 2/\bar{\rho}$ is the radius of the circular Brillouin zone of the reciprocal space.

The thickness-dependent critical density for the electron WC effect can be found for our case from its PF model analogue using equation (S32) as follows

$$\Gamma_0 = 2.8 = \frac{e^2/\bar{\rho}}{\hbar^2 \pi N_{2D}^c/(2m^*)} = \frac{e^2/\sqrt{\pi N_{2D}^c}}{\hbar^2 \pi N_{2D}^c/(2m^*)} \rightarrow \frac{1}{\bar{\varepsilon}(2\tilde{\varepsilon}d\sqrt{\pi N_{2D}^c}+1)} \frac{e^2/\sqrt{\pi N_{2D}^c}}{\hbar^2 \pi N_{2D}^c/(2m^*)}$$

This leads to the quadratic equation with only one root being positive. In dimensionless variables after simplifications one obtains

$$n_c(d) = \frac{1}{\pi \Gamma_0 \bar{\varepsilon}\tilde{\varepsilon}\tilde{d}} \frac{\sqrt{1+16\tilde{\varepsilon}\tilde{d}/(\Gamma_0\bar{\varepsilon})}-1}{\sqrt{1+16\tilde{\varepsilon}\tilde{d}/(\Gamma_0\bar{\varepsilon})}+1} \tag{S33}$$

Comparing the theoretical $N_{2D}^c$ with the experimental values in Table S8 shows that the 2 nm thick film belongs to the Wigner crystal domain. Although, our theoretical calculations suggest the possibility of Wigner crystallization in 2 nm ultrathin films, but direct experimental evidence of electron crystal in TD materials can be explored as a future research.

**Section V. Computational details for the simulation of optical properties from first-principle**

First-principle density functional theory (DFT) and density functional perturbation theory (DFPT) calculations were used further to explain the role of electronic structure, phonons, and electron-phonon coupling on the optical response of HfN and to model the dielectric response, using the plane wave basis pseudopotential method, as implemented in the Quantum ESPRESSO (QE) package [56]. The Perdew, Bruke, and Ernzerhof (PBE) parametrization of generalized gradient approximation (GGA) exchange-correlation functional was used for the calculations [57]. Several pseudization schemes were taken for different types of calculations such as ultra-soft Vanderbilt pseudopotential (USVBPP) for linear-response time-dependent-DFPT (TD-DFPT) calculation [58], norm-conserving (NC) Troullier-Martins (TM) pseudopotential scheme for complex dielectric function simulation using independent particle (IP) approximation [59], and ONCV NC pseudopotential for the electron-phonon coupling (EPC) parameter calculations [60]. An energy cutoff (ENCUT) of 100 Ry for NC pseudopotentials and 50 Ry for USVBPP was used. The charge density cutoff was limited to four times the ENCUT for NC pseudopotential and 10 times for the USVBPP, except for the phonon calculation, for which the charge density cutoff was restricted to at least 8 times of ENCUT to avoid phonon instability. The Brillouin-zone (BZ) was sampled by the Monkhrost-Pack scheme with an unshifted k-point grid of 28×28×28 for bulk and 50×50×1 for the slab models with the Marzari-Vanderbilt cold smearing of 0.002 Ry [61,62]. The optimized lattice constant of bulk HfN was found to be 4.54 Å from the DFT calculation. For the modelling of ultrathin HfN (001) film containing 2 and 10 layers, a slab model was considered with an in-plane lattice constant same as the bulk HfN, along with a 17 Å vacuum to avoid any spurious interactions between the periodic replicas. Each slab was fully relaxed until the average forces on the individual atom became less than 0.026 eV/Å.

The complex dielectric response in the main text was simulated using the TD-DFPT approach as implemented in the TURBOEELS code in the QE suite [49]. The code implements a Liouville-Lanczos approach to simulate the dielectric response of solid under linear-response theory. Under the linear-response approximation, the dielectric response of the material $\varepsilon(\mathbf{q}, \omega) = \varepsilon' + i\varepsilon''$ depends on the transferred momentum $\mathbf{q}$ and frequency $\omega$. Both the crystal local field effects and exchange-correlation local field effects were considered in the TD-DFPT calculation [63]. The dielectric response was calculated with a small momentum transfer (~0.028 Å$^{-1}$) in the Γ-X direction of the BZ with a Lorentzian broadening of 0.02 Ry for charge-density susceptibility. The converge spectrum has been obtained with an initial 8000 Lanczos iterations along with 30,000 iteration steps for the extrapolation of Lanczos coefficients. The real and imaginary parts of the dielectric response function show smooth spectra in agreement with the experiment without any spurious wiggles, which conforms to the proper convergence of the spectra [64]. The optical response in the vanishing momentum limit ($\mathbf{q} \to 0$) was calculated in the random phase approximation (RPA) without including the local field effects using the SIMPLE code at exactly $\mathbf{q} = 0$ [65]. The code utilizes the IP formalism of the frequency-dependent Drude-Lorentz model with optimal product basis (OPB). The expression of the IP dielectric constant for metal with negligible transferred momentum of a photon is described as a sum of intra-band Drude-like term due to the conduction electron at the Fermi surface and inter-band Lorentz-like term due to the transition between occupied and unoccupied electronic states.

$$\varepsilon_{IP}(\hat{\mathbf{q}}, \omega) = \varepsilon_{IP}^{inter}(\hat{\mathbf{q}}, \omega) + \varepsilon_{IP}^{intra}(\hat{\mathbf{q}}, \omega) \tag{S34}$$

where,

$$\varepsilon_{IP}^{inter}(\hat{\mathbf{q}}, \omega) = 1 - \frac{4\pi}{V} \sum_{\mathbf{k}} \sum_{n \neq n'} \frac{|\langle \Psi_{n'\mathbf{k}} | \hat{\mathbf{q}} \cdot \mathbf{v} | \Psi_{n\mathbf{k}} \rangle|^2}{(E_{n'\mathbf{k}} - E_{n\mathbf{k}})^2} \frac{f_{n\mathbf{k}} - f_{n'\mathbf{k}}}{\omega - (E_{n'\mathbf{k}} - E_{n\mathbf{k}}) + i\eta} \tag{S35}$$

$$\varepsilon_{IP}^{intra}(\hat{\mathbf{q}}, \omega) = -\frac{\omega_P^2(\hat{\mathbf{q}})}{\omega(\omega+i\gamma_D)}. \tag{S36}$$

The Drude-plasma frequency is defined as

$$\omega_P^2(\hat{\mathbf{q}}) = \frac{4\pi}{V}\sum_n\sum_{\mathbf{k}}|<\Psi_{n\mathbf{k}}|\hat{\mathbf{q}}\cdot\mathbf{v}|\Psi_{n\mathbf{k}}>|^2\left(-\frac{\partial f}{\partial E}\right). \tag{S37}$$

Where, $\mathbf{v}$ is the velocity operator, $\eta$ and $\gamma_D$ are empirical broadening parameters corresponding to the inter- and intra-band scattering, respectively. V is the volume of the crystal and $f_{n\mathbf{k}}$ is the Fermi-Dirac distribution function [65]. The OPB was constructed using 6×6×6 k-mesh of nscf calculation with a threshold of 0.01 bohr. We have calculated the response function using a large broadening of 0.01 Ry for Drude plasma frequency and 0.01Ry broadening for both $\eta$ and $\gamma_D$. We have checked that the dielectric response obtained for $|\mathbf{q}| \leq 0.028 \text{ Å}^{-1}$ from TD-DFPT is in good agreement with the IP formalism of the Drude-Lorentz model. Although the real part of the complex dielectric function matches very well with almost the same ENZ wavelength for both schemes (see Fig. S6), a small deviation in the imaginary part at lower energy is due to the use of different pseudopotentials in the two calculations and negligible local field effects in the calculation with SIMPLE code at $\mathbf{q} = 0$ [66].

The electronic band structure and density of states calculations show HfN to be a metal with the Fermi level crossing the states mostly dominated by Hf-*5d* orbitals (see Fig. 4b and Fig. S7a). The density of states (DOS) shows three separate sets of bands, each with a dominant contribution from the single element and orbital (ionic contribution) with a minor contribution from other elements (covalent contribution). The band at higher energy (peak at ~ -16 eV) is mostly dominated by N-*2s* character, marked as $a_1$. The band with index $a_2$ from -9.0 eV to -3.0 eV is composed of mainly N-*2p* characters along with a small hybridization from Hf-*5d* orbitals. The electronic band structure (see Fig. S7a) shows that $a_2$ is mainly contributed from three dispersive bands of N-*2p* character and the sharp peak at (~ -6 eV) originates from the flat bands for some portion in the BZ (X-W-L). The $a_3$ band is mostly dominated by a bundle of five dispersive bands with Hf $t_{2g}$ character with a small hybridization from N-*2p* character as shown in Fig. S7b. These bands are responsible for the metallicity of HfN. Among these bands, the three lowest energy bands degenerate at Γ and cross the Fermi level at several points of BZ, suggesting a multi-electron pocket Fermi surface of HfN (see Fig. S8) that gives rise to strong intra- and inter-band scattering channel for HfN as compared to the other good metals like Au, Ag in which the Fermi level crosses only the *s*-bands with almost ideal free electron behaviour [67,68]. This suggests that the EPC will be higher in transition metal nitrides (TMN) (e.g., TiN, HfN) as compared to the elemental *s*-band metals, which leads to a higher damping coefficient in TMN. To quantify this, we have explicitly calculated the EPC constant (λ) from Migdal-Eliashberg theory for two TMN TiN and HfN and compared it with the EPC constant of Au and Ag for better understanding.

The phonon dispersion of both TiN and HfN shows acoustic phonon mode softening at q ~ 0.7 along the Γ-to-X direction and Γ-to-K direction due to the nesting of the Fermi surface by wave vector q near the zone centre along the above-specified directions (see Fig. S8 and Fig. S9), consistent with the previous literature reports [69–71]. Both HfN and TiN show large phononic bandgap due to the mass difference of transition metals and nitrogen. Due to the large mass difference between Ti and Hf, acoustic phonon energy is less in HfN as compared with TiN with a phononic bandgap of ~ 40 meV (see Fig. S9). The acoustic mode contributions are coming from Ti and Hf atoms and the optical mode contribution is mainly due to the N atom vibrations. The acoustic modes of TMN mostly contribute to the EPC as shown in Fig. S10. The cumulative frequency-dependent EPC λ(ω) shows (see Fig. S10) a much higher value of 0.72 compared to other Drude metals and TiN (4.5 times than Ag, 3.3 times than Au, and 1.3 times than TiN (see Table S9)), which suggests a strong EPC in HfN. The calculated superconducting critical temperature ($T_c$) matches well the experimental $T_c$ value for our calculated EPC constant $\lambda$ [31]. There are also several other factors like dislocations, grain boundary, and impurity scatterings associated with the damping constant that we have not considered in our theoretical modelling. Although it can be qualitatively described that the grain boundary resistance is higher in HfN due to the anisotropic nature of the Fermi surface ( see Fig. S8) as compared with elemental metals (such as Au and Ag), where the Fermi surface is nearly isotropic with spherical geometry [72]. Such high $\lambda$, along with large dislocations scattering arising from ~ 7.3% lattice-mismatch in HfN films deposited on (001) MgO substrates, grain boundary, and impurity scatterings are the main intrinsic reasons behind its large $\gamma_D$, which correlates with the experimental observations.

Calculated inter-band contribution to the dielectric function from the SIMPLE code shows a strong contribution of intra-band scattering responsible for the plasmonic contribution of HfN. Above ~ 3.5 eV, the contribution is mainly from the inter-band contribution due to the N-*2p* to Hf-*5d* transition as shown in Fig. S11. The small peak in the real part of the dielectric constant is due to the inter-band transition as seen from the presence of a peak at the same position (~ 250 nm) in the inter-band contribution of the real part. Also, the peak in the imaginary part of the dielectric constant at ~ 200 nm (~ 6.2 eV) corresponds to the inter-band transition between N-*2p* to Hf-*5d* states at the Γ point. Experimental data also show a similar peak in the imaginary part of dielectric constant near ~ 220 nm (see Fig. 1D in the main text). The imaginary part ($\varepsilon''$) diverges near $\omega \to 0$, which corroborates with the finite DC conductivity of metals [44]. The real part ($\varepsilon'$) changes sign from negative-to-positive with a positive slope in the frequency scale with minimum $\varepsilon''$ indicates the ENZ frequency [65]. The ENZ frequency of HfN is in the UV range of the electromagnetic spectrum suggesting that light with a frequency higher than ENZ frequency is not perfectly reflecting. $\varepsilon'$ becomes positive in the range [3.6, 8.8] eV (see Fig. S12a and Fig. S12b) due to the inter-band transition as described above. Calculated reflection (R) and transmission (T) coefficients (see Fig. S12c and Fig. S12d) show that HfN is fully reflecting (> 90%) in the IR and some portion of the visible range above (~ 560 nm), which corresponds to the yellow-brownish colour of HfN. The reflection spectrum shows a dip near ~ 325 nm is in good agreement with previous experimental results. However, at higher energies (4-10 eV) T does not tend to unity which implies that HfN is not an ideal free electron metal with a partial lossy character and is not fully transparent to the UV region of the electromagnetic spectrum. To further validate this we have calculated the electron energy loss spectrum (EELS) $L = -Im[\varepsilon^{-1}(\mathbf{q}, \omega)]$, depends on the transferred momentum $\mathbf{q}$ and frequency $\omega$ [49,64]. Simulated EELS of bulk HfN shows peak at 3.5 eV (see Fig. S13), the zero cross-over point from negative to positive $\varepsilon'$ in the frequency scale corresponds to the screened plasmon. The screened plasmon peak broadens and quenches in intensity, along with a positive dispersion as the transferred momentum $|\mathbf{q}|$ increases. With the higher momentum transfer a new feature marked as $E_{el-h}$ arises in the lower energy region of the spectrum. This can be parametrized by critical momentum $q_c$. The value of $q_c$ for HfN is 0.41 Å$^{-1}$ above which, the plasmon oscillations are quenched due to their transfer of energy to the band electrons undergoing interband transition within the multiple Hf-*5d* bands crossing the Fermi level. The value of $q_c$ for HfN is lower than that of TiN reported as 0.59 Å$^{-1}$ [44]. Plasmon decay to the electron-hole pair in the lower critical momentum region suggests the possibility of strong interband transition with higher loss. This further validates HfN as not an ideal free-electron metal with the possibility to explore experimentally the transdimensional plasmonic nature of HfN.

In literature, bulk HfN ENZ wavelength varies from 350 nm to 430 nm [6,7]. We have addressed this in terms of the presence of Hf vacancy, which is possible to form according to the stable formation energy from the previous DFT calculation [73]. Our calculated dielectric spectra from both TD-DFPT and IP formalism do not match well with the present experimental scenario. With possible Hf vacancy or reduced electron concentration in the system, we have arrived at a good agreement with the experimental results (TD-DFPT result in Fig. S14). The ENZ wavelength of pristine HfN is found at 357 nm in the modelling which changes to 435 nm after incorporating Hf vacancy in the calculation. Depending on the growth condition, as-deposited HfN films contain different defect concentrations as a result, carrier concentration and ENZ wavelength changes. We have listed some experimental ENZ wavelengths of HfN films available in the literature and compared them with our present HfN film (see Table S10). The presence of such vacancy explains a slightly higher ENZ wavelength in HfN compared to the previous reports. Although the calculated $\varepsilon'$ matches quite well with the experimental observation, the small mismatch of the $\varepsilon''$ at longer wavelengths (IR region) is due to the smaller Drude damping in modelling compared to the experimental case. We have simulated the role of defects by reducing the electron concentration (see tot_charge tag of pw.x executable) from the unit cell used in the DFT calculation while considering a compensating jellium background to avoid possible divergences. Also, the calculated Drude plasma frequency ($\omega_P$) for the pristine HfN from the SIMPLE code is 9.0 eV, slightly higher than the present experimental value of 7.85 eV. After reducing the electrons from the unit cell $\omega_P$ turns out to be 8.32 eV, close to the experimental value of the bulk HfN (see Table S11). This explains the role of Hf vacancy in controlling the ENZ wavelength of HfN films.

Finally, we have calculated the electronic density of states and optical response of bilayer HfN and compared it with the bulk to provide significant insight. As presented in Fig. 4c, the valence bandwidth and distance of the valence band tail from $E_F$ are lower in thinner films than in the bulk HfN. Since localization generally sets in from the band tails, the narrower bands close to the $E_F$ should facilitate the electronic confinement. Finally, we have simulated the optical response for the ultrathin HfN films using IP approximation. The red shift of the ENZ wavelength is there (see Fig. S15 and Table S12), consistent with the previous modelling result of ultrathin TiN films and recent electromagnetic theory for describing confinement-induced optical properties of ultrathin plasmonic metals [74]. This suggests the picture of confinement of electrons in ultrathin HfN films, which dominates over other contributions to make the ultrathin HfN films become completely dielectric. Although first-principle DFT cannot capture these localization effects [46], which needs other high-level computation schemes or model Hamiltonian approach to properly describe the theoretical origin of the localization in ultrathin HfN films.

**Supplementary Figures**

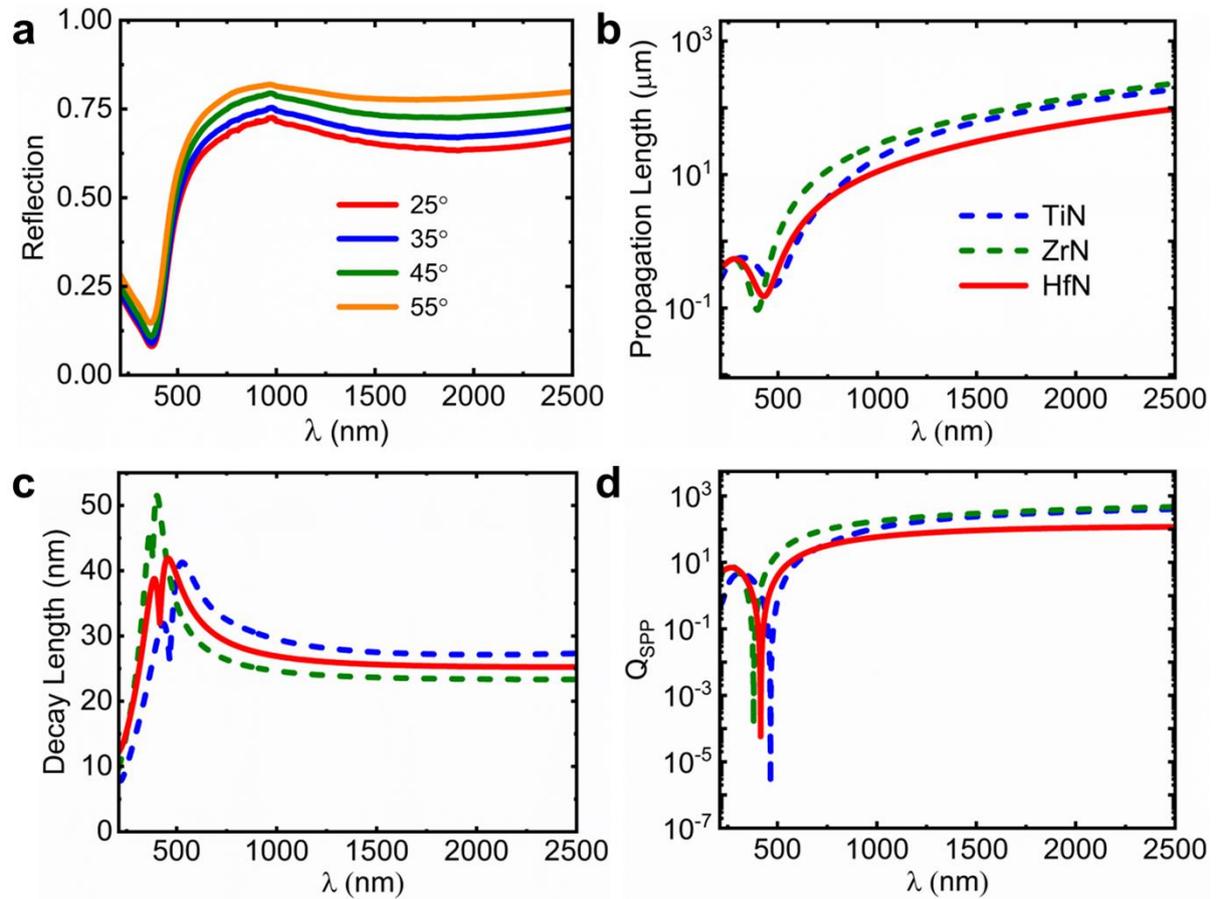

**Fig. S1 | Surface plasmon polariton parameters of bulk HfN.** (**a**) Angle-dependent reflection spectra of bulk HfN film. Comparison of (**b**) propagation length, (**c**) decay length, and (**d**) quality factor of bulk HfN film with TiN and ZrN films.

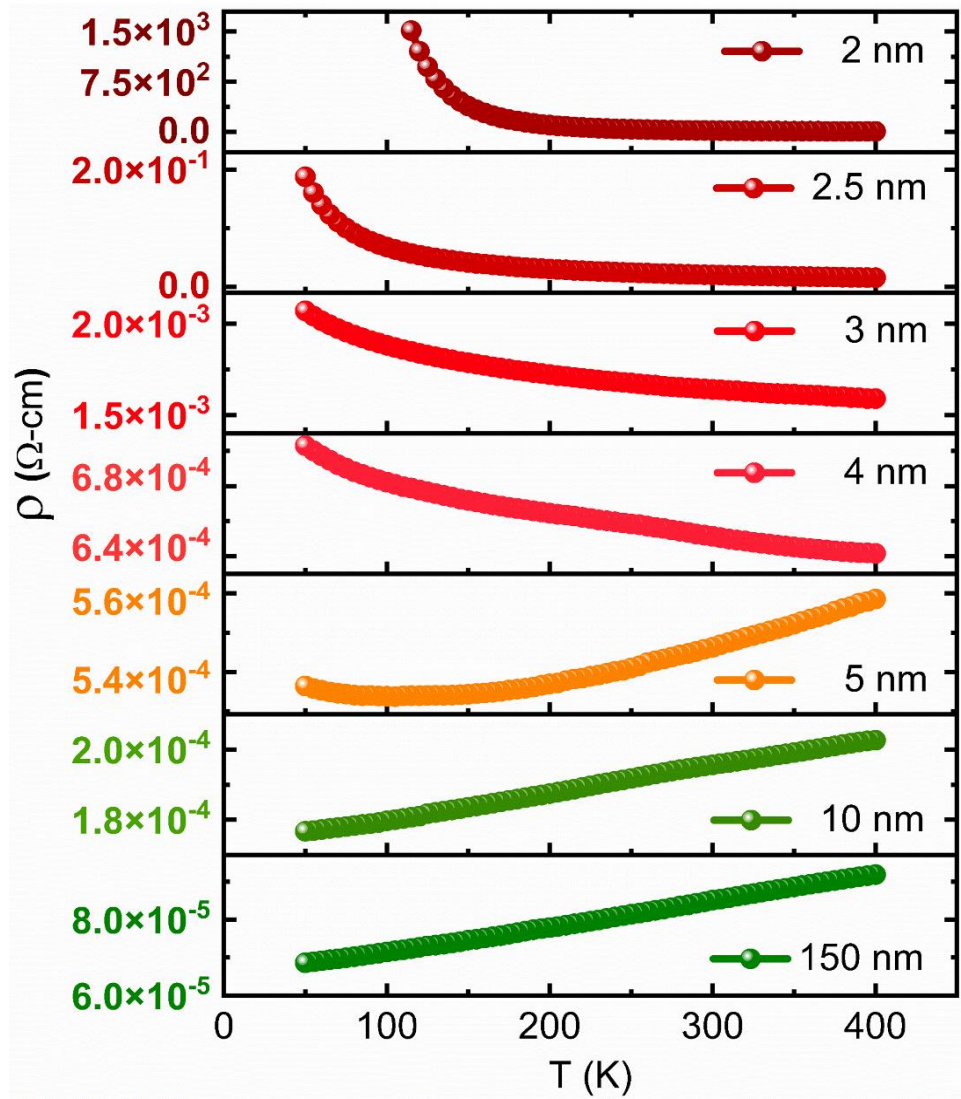

**Fig. S2 | Temperature-dependent resistivity of HfN film with different thickness.**

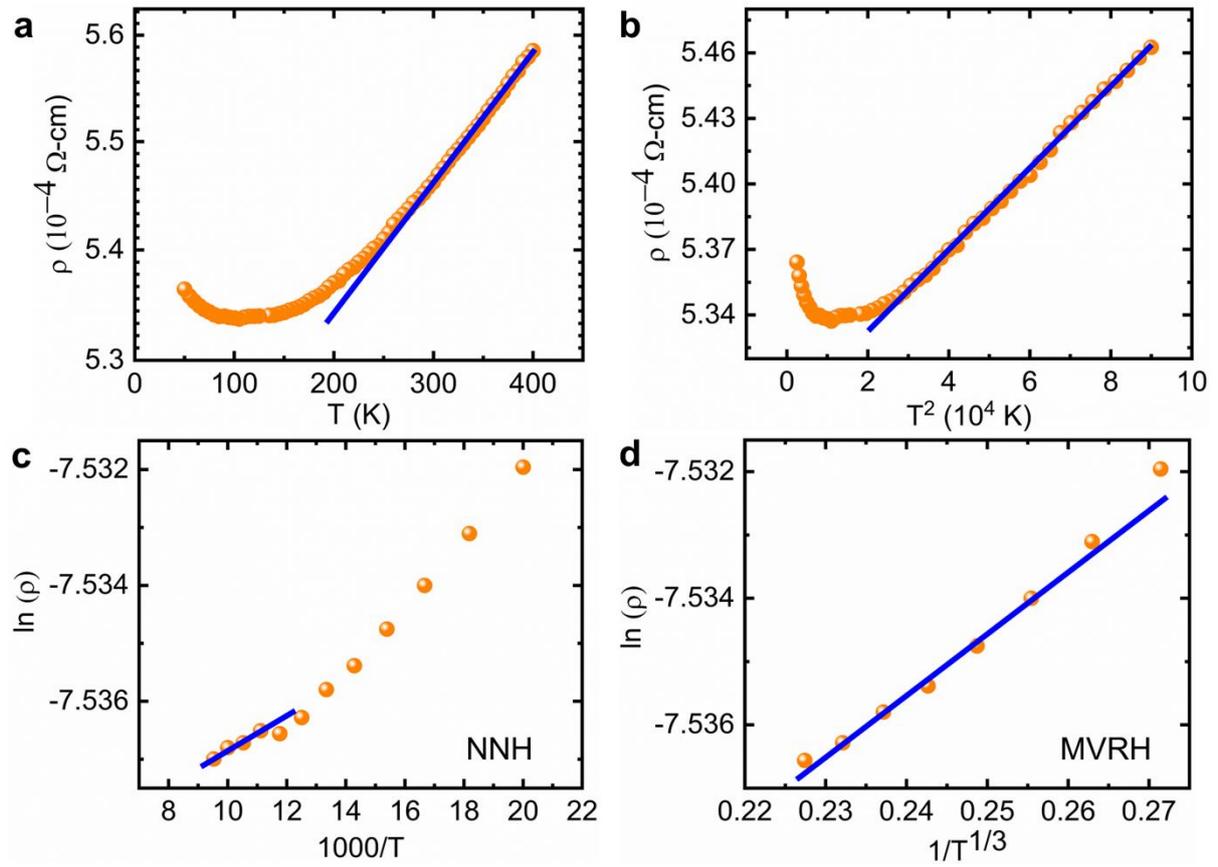

**Fig. S3 | Temperature-dependent resistivity plots of 5 nm HfN film fitted for different mechanisms.** (**a**) Pure metallic ($\propto T$), (**b**) $\propto T^2$, (**c**) nearest neighbour hopping (NNH), and (**d**) Mott variable range hopping (MVRH).

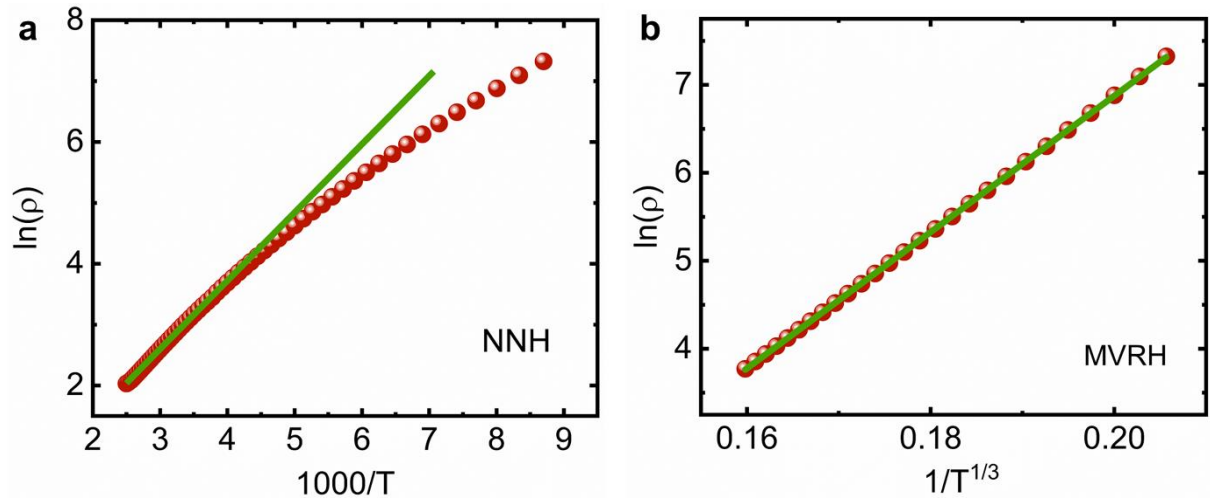

**Fig. S4 | Temperature-dependent resistivity plots of 2 nm HfN film fitted for different mechanisms.** (**a**) Nearest neighbour hopping (NNH) and (**b**) Mott variable range hopping (MVRH).

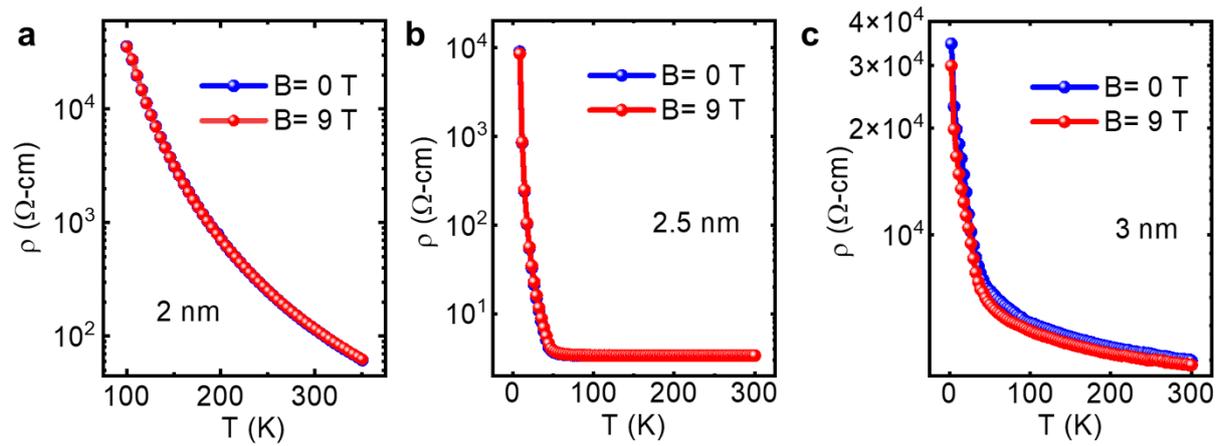

**Fig. S5 | Resistivity plots of different HfN films in presence of magnetic field. (a)** 2 nm, (**b**) 2.5 nm, and (**c**) 3 nm HfN films with and without magnetic field.

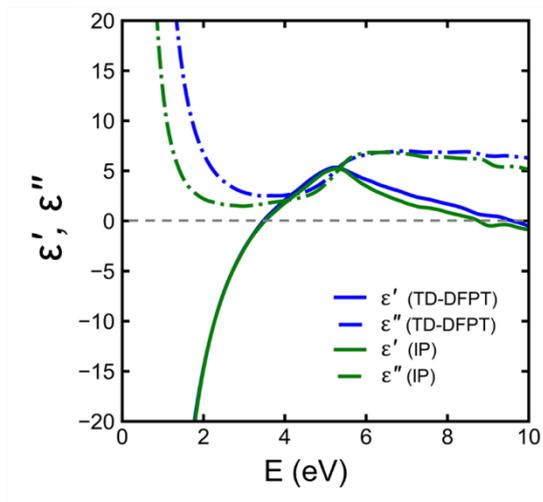

**Fig. S6 | Real and imaginary parts of the dielectric response of bulk HfN from TD-DFPT and IP approximation.**

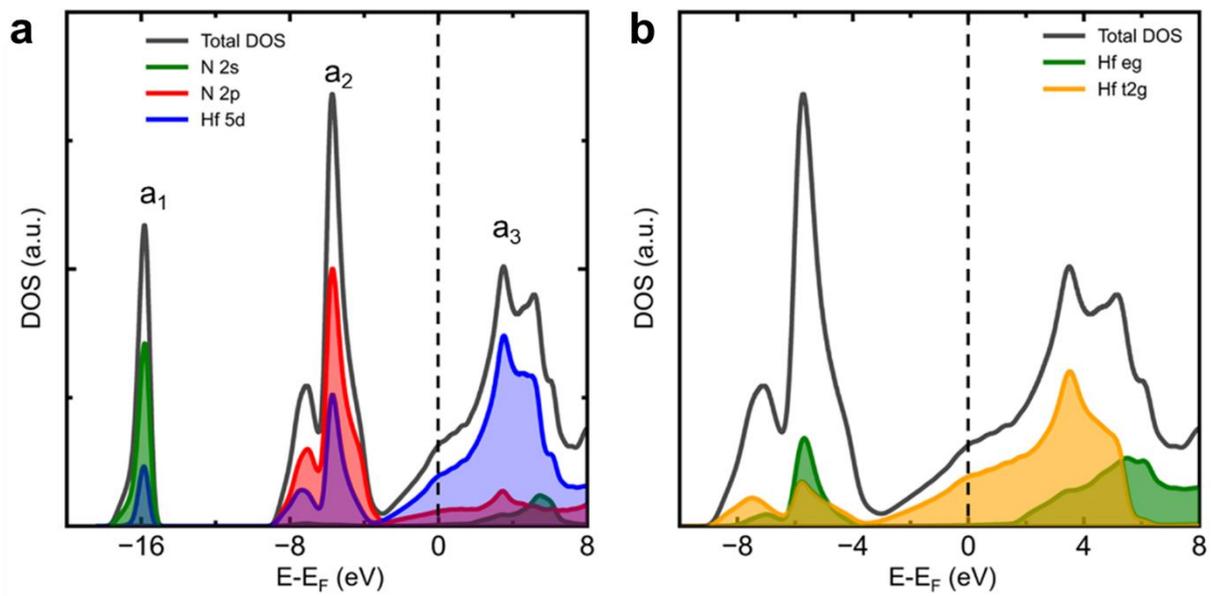

**Fig. S7 | Electronic DOS of bulk HfN.** (**a**) Total DOS (black line), N-*2s* projected DOS (green shaded area), N-*2p* projected DOS (red shaded area), and Hf-*5d* projected DOS (blue shaded area). (**b**) $e_g$-$t_{2g}$ splitting of Hf-*5d* orbital. $e_g$ orbital (green shaded area) and $t_{2g}$ orbital (orange shaded area).

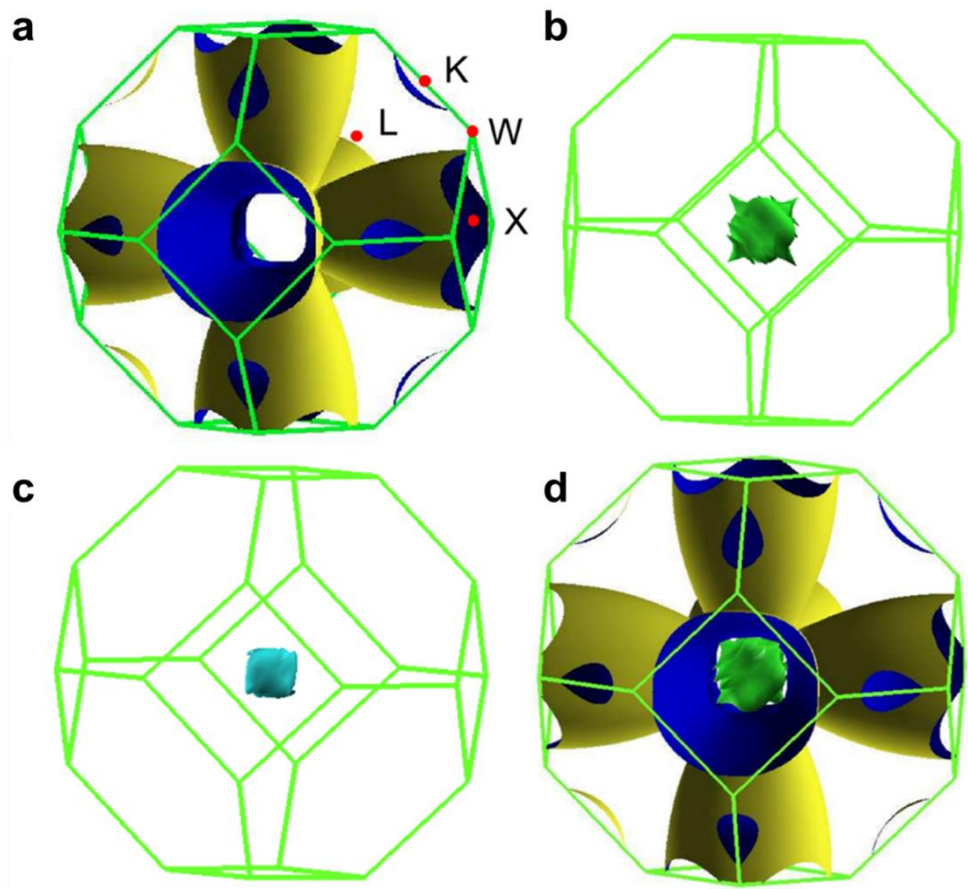

**Fig. S8 | Different branches of the Fermi surface of HfN.** (**a**) Fermi surface corresponds to the first band. (**b**) Fermi surface corresponds to the second band. (**c**) Fermi surface corresponds to the third band. (**d**) Fermi surface corresponds to the three merged bands together.

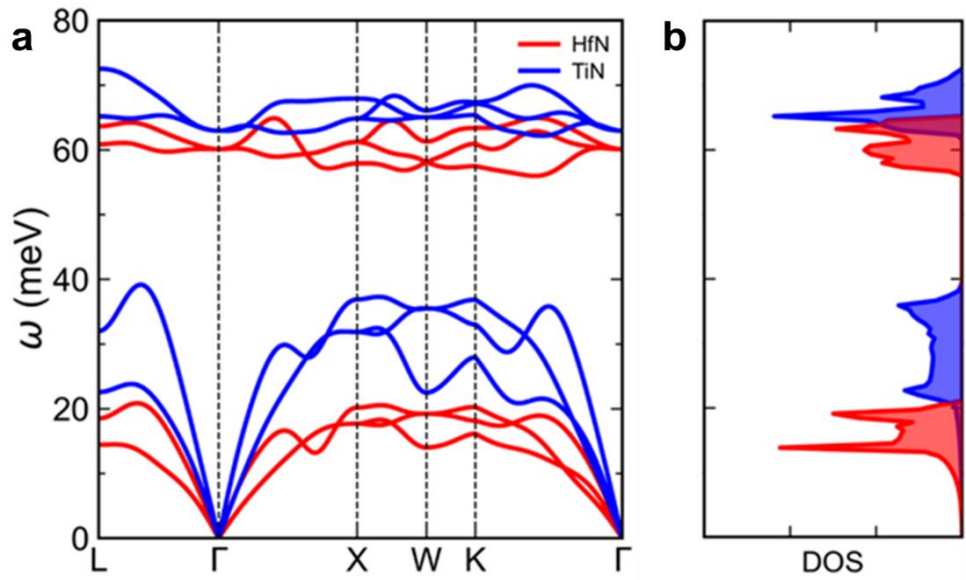

**Fig. S9 | (a) Phonon dispersion of two transition metal nitrides, TiN and HfN. (b) Phonon density of states of TiN and HfN. TiN dispersion and DOS (blue colour) and HfN dispersion and DOS (red colour).**

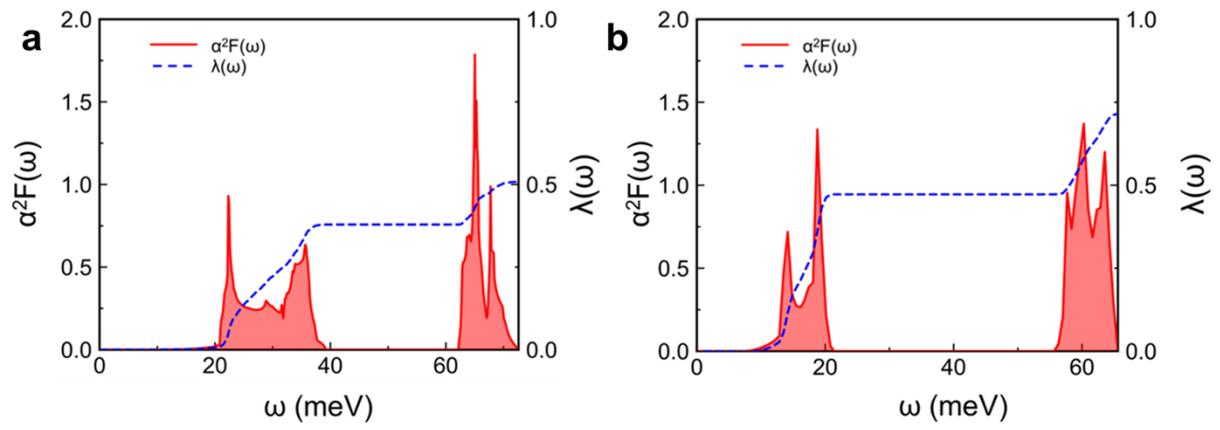

**Fig. S10 | Eliashberg spectral function $\alpha^2F(\omega)$ and cumulative frequency dependent EPC constant $\lambda(\omega)$.** (a) TiN and (b) HfN.

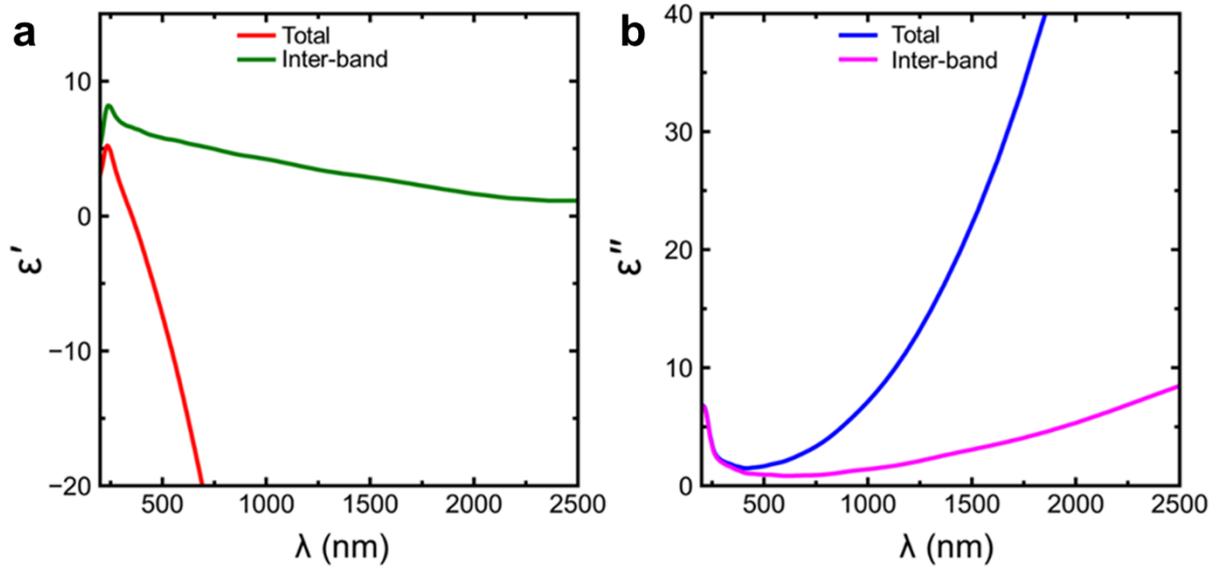

**Fig. S11 | Contribution of inter-band transition to dielectric permittivity.** (a) real ($\varepsilon'$) and (b) imaginary ($\varepsilon''$) parts of the dielectric permittivity of HfN.

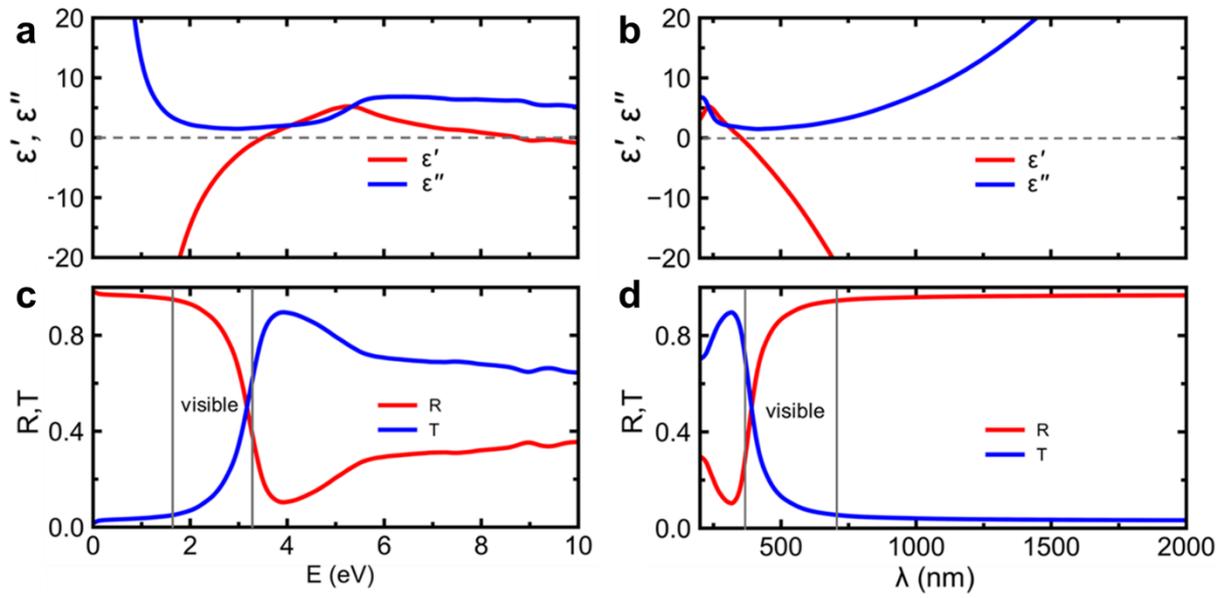

**Fig. S12 | Theoretically calculated optical properties of HfN.** The real and imaginary parts of the dielectric constant of HfN in (**a**) energy (eV) scale and (**b**) wavelength (nm) scale calculated using the SIMPLE code. Calculated reflectivity (R) and transmissivity (T) of HfN in (**c**) energy (eV) scale and (**d**) wavelength (nm) scale using the Fresnel equation at normal incident conditions.

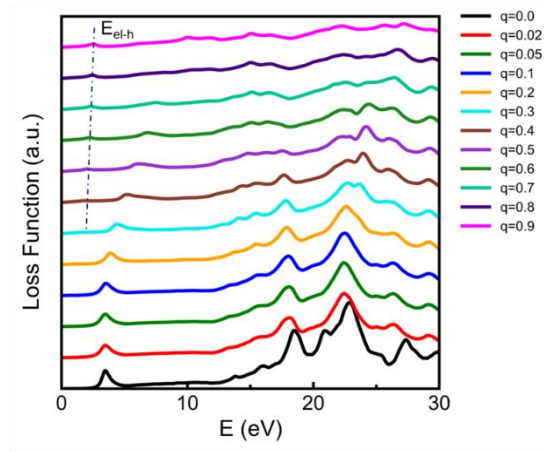

**Fig. S13 | Theoretically calculated EELS of bulk HfN with increased transferred momentum |q| in units of (2π/A).** The electron-hole (el-h) contribution arises above the critical momentum transfer as marked by the dot-dash line below the screened plasmon energy.

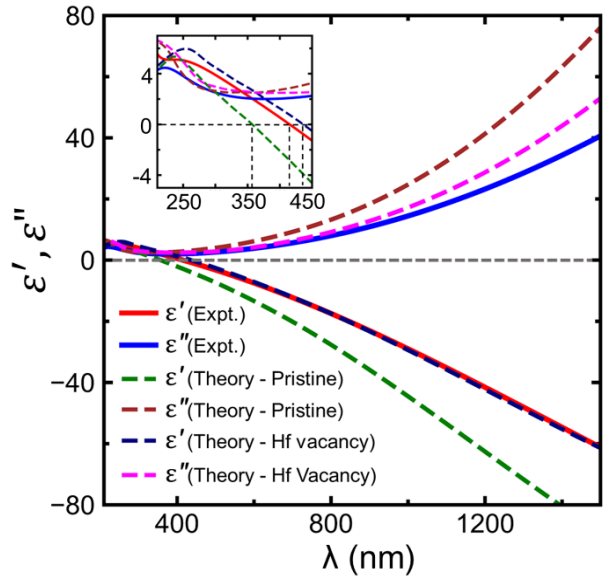

**Fig. S14 |** Experimental and theoretical $\varepsilon'$ and $\varepsilon''$ of bulk HfN with and without Hf vacancy are presented. The inset highlighs the ENZ region.

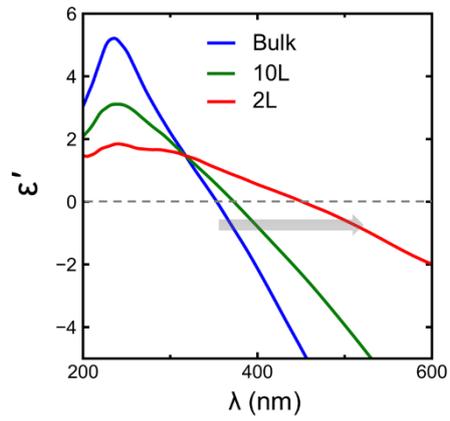

**Fig. S15 | The real part of the dielectric constant of bulk (blue) and ultrathin HfN film having 10 layers (green) and 2 layers (red) with a red shift of the ENZ wavelength going from bulk to the ultrathin dimension.**

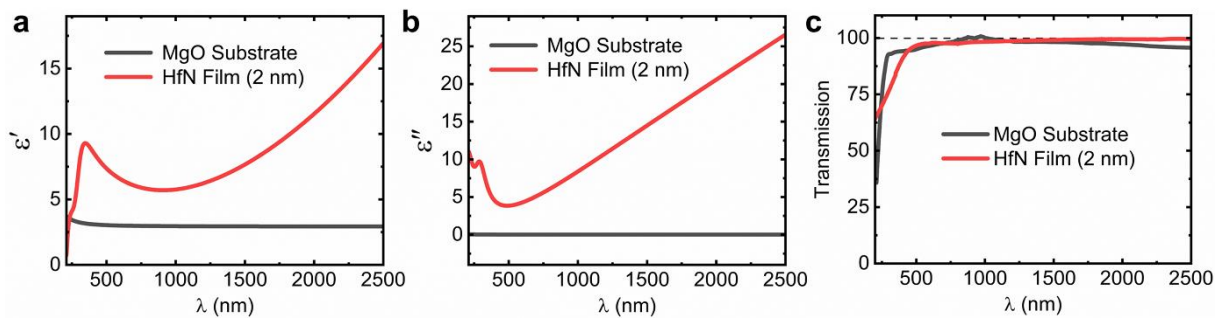

**Fig. S16 | Comparison of Optical Properties of 2 nm HfN film and MgO Substrate**. The real (**a**) and imaginary (**b**) parts of the dielectric permittivity of the MgO substrate and a 2 nm HfN film are presented to highlight the distinct optical properties of each material. The dielectric constant of MgO remains nearly constant at approximately 3 across the entire UV-Visible-Near-IR spectral range. In contrast, the real part of the permittivity of the 2 nm HfN film exhibits characteristics similar to those of thicker HfN films (Fig. **1e**). (**c**) In the Near-IR spectral range, the optical transmission of the 2 nm HfN film is almost identical to that of the MgO substrate. However, there are notable differences in the UV-Visible range. When performing the transmission spectrum measurements, the effect of the substrate is accounted for by conducting baseline measurements. Although the confinement effect is influenced by the dielectric constant of the substrate, the optical properties observed in the 2 nm HfN film cannot be attributed to the influence of the transparent MgO substrate alone. This indicates that the optical properties of the 2 nm HfN film arise from its inherent characteristics rather than being dominated by the MgO substrate.

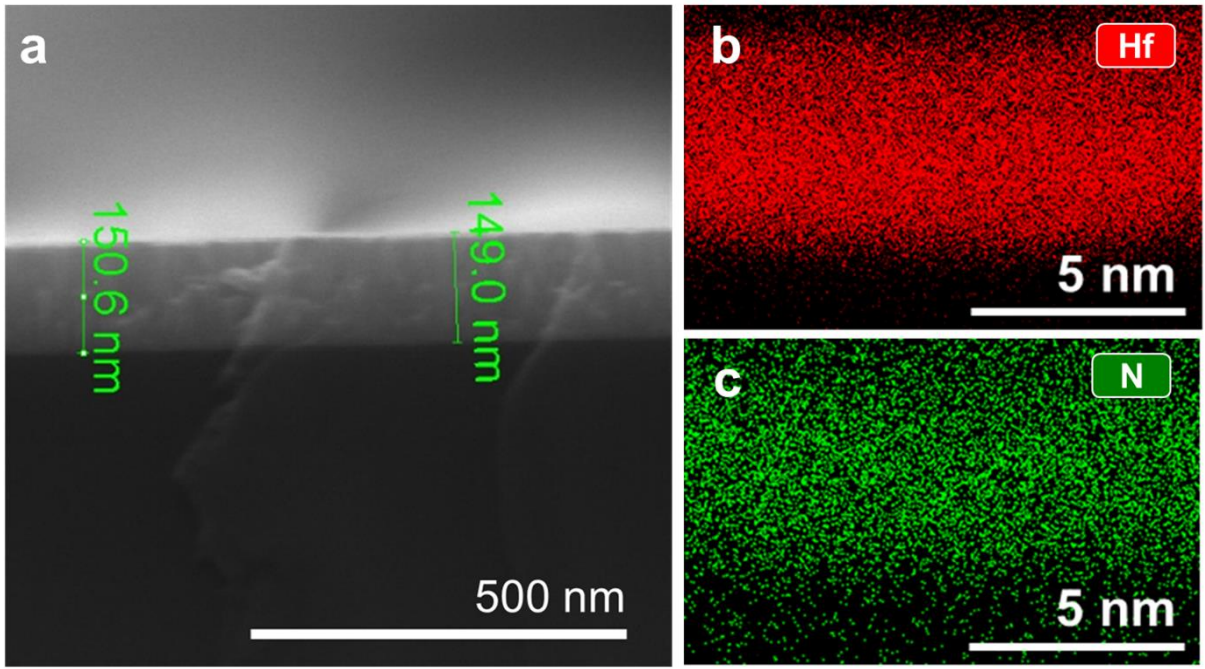

**Fig. S17 | Cross-sectional SEM & EDS maps**. (**a**) Cross-sectional field-emission scanning electron microscopy (FESEM) image of bulk HfN film. The thickness of the bulk HfN film is found to be 150 nm. For ultrathin films, the thickness is measured by fitting the X-ray reflectivity spectra (XRR). Transmission electron microscopy (TEM)-Energy-dispersive X-ray spectroscopy (EDS) elemental maps (raw data) of (**b**) Hf and (**c**) N show uniform homogeneous distribution within the film.

# Supplementary Tables

**Table S1: Optical parameters (Drude) obtained from the ellipsometry data fitting for various HfN films.**

| Thickness (nm) | $\varepsilon_{core}$ | $\omega_P$ (eV) | $\gamma_D$ (eV) | $\lambda_{ENZ}$ (nm) | $\varepsilon^{"} @ \lambda_{ENZ}$ |
|---|---|---|---|---|---|
| 2 | 7.15 | 5.08 | 1.84 | ------ | ------ |
| 2.5 | 6.33 | 5.36 | 1.48 | ------ | ------ |
| 3 | 5.84 | 5.45 | 1.35 | ------ | ------ |
| 4 | 5.75 | 5.68 | 1.33 | 787 | 6.75 |
| 5 | 4.7 | 5.96 | 1.05 | 630 | 4.44 |
| 7.5 | 4.03 | 6.5 | 0.93 | 535 | 3.47 |
| 10 | 2.93 | 6.76 | 0.86 | 503 | 3.15 |
| 15 | 2.85 | 7.16 | 0.78 | 478 | 2.87 |
| 150 | 2.82 | 7.85 | 0.52 | 416 | 2.08 |

**Table S2: Optical parameters (Drude and Lorentz) obtained from ellipsometry data fitting for 150 nm film.**

| T (K) | $\varepsilon_{core}$ | $\omega_P$ (eV) | $\gamma_D$ (eV) | $f_1$ | $\gamma_1$ (eV) | $\omega_{0,1}$ (eV) |
|---|---|---|---|---|---|---|
| 200 | 2.1 | 7.93 | 0.49 | 4.47 | 2.9 | 5.8 |
| 300 | 2.56 | 8.06 | 0.52 | 4.22 | 2.91 | 5.71 |
| 400 | 3.07 | 8.15 | 0.55 | 4.48 | 2.83 | 5.69 |
| 500 | 2.54 | 8.19 | 0.58 | 4.23 | 2.84 | 5.55 |
| 600 | 2.86 | 8.24 | 0.62 | 4.50 | 2.85 | 5.63 |
| 700 | 1.7 | 9.05 | 0.67 | 2.96 | 2.42 | 5.02 |

**Table S3: Optical parameters (Drude and Lorentz) obtained from ellipsometry data fitting for 5 nm film.**

| T (K) | $\varepsilon_{core}$ | $\omega_P$ (eV) | $\gamma_D$ (eV) | $f_1$ | $\gamma_1$ (eV) | $\omega_{0,1}$ (eV) | $f_2$ | $\gamma_2$ (eV) | $\omega_{0,2}$ (eV) |
|---|---|---|---|---|---|---|---|---|---|
| 200 | 5.64 | 5.73 | 1.06 | 10.79 | 2.3 | 7.13 | 3.68 | 1.78 | 4.5 |
| 300 | 6.31 | 5.74 | 1.08 | 6.83 | 2.31 | 6.2 | 3.16 | 1.5 | 4.3 |
| 400 | 6.12 | 5.74 | 1.13 | 6.4 | 2.45 | 6.1 | 2.94 | 1.4 | 4.24 |
| 500 | 5.6 | 5.74 | 1.16 | 6.33 | 2.52 | 6.23 | 3.2 | 1.56 | 4.31 |
| 600 | 5.73 | 5.75 | 1.2 | 7.56 | 2.42 | 6.52 | 3.42 | 1.7 | 4.39 |
| 700 | 4.11 | 5.75 | 1.26 | 6.64 | 2.36 | 6.02 | 3.05 | 1.44 | 4.2 |

**Table S4: Optical parameters (Drude and Lorentz) obtained from ellipsometry data fitting for 2 nm film.**

| T (K) | $\varepsilon_{core}$ | $\omega_P$ (eV) | $\gamma_D$ (eV) | $f_1$ | $\gamma_1$ (eV) | $\omega_{0,1}$ (eV) | $f_2$ | $\gamma_2$ (eV) | $\omega_{0,2}$ (eV) |
|---|---|---|---|---|---|---|---|---|---|
| 200 | 7.35 | 4.15 | 1.85 | 8.98 | 2.56 | 6.05 | 5.32 | 1.62 | 4.26 |
| 300 | 7.45 | 4.1 | 1.93 | 9.52 | 2.58 | 6.38 | 5.54 | 1.68 | 4.29 |
| 400 | 7.34 | 4.05 | 1.98 | 9.03 | 2.37 | 6.3 | 5.9 | 1.84 | 4.34 |
| 500 | 6.85 | 3.97 | 2.02 | 8.7 | 2.61 | 5.92 | 4.86 | 1.49 | 4.11 |
| 600 | 7.15 | 3.63 | 2.05 | 9.21 | 2.72 | 6.1 | 5.02 | 1.65 | 4.2 |
| 700 | 7.65 | 3.46 | 2.12 | 9.01 | 2.46 | 5.74 | 4.57 | 1.44 | 3.99 |

**Table S5: Fitting parameters extracted from the MVRH model.**

| Thickness (nm) | $T_M$ (K) | $E_a^M$ (eV) |
|---|---|---|
| 2 | $4.72 \times 10^6$ | 400.71 |
| 2.5 | $5.98 \times 10^4$ | 0.5157 |
| 3 | 6.37 | $5.5 \times 10^{-4}$ |
| 4 | 0.185 | $1.6 \times 10^{-5}$ |
| 5 | $7.69 \times 10^{-4}$ | $6.63 \times 10^{-8}$ |

**Table S6: Temperature coefficient of resistivity of different metals.**

| Metal | Temperature coefficient of resistivity ($10^{-3}$/°C) |
|---|---|
| Ag | 3.8 |
| Au | 3.4 |
| Cu | 3.86 |
| Al | 4.29 |
| W | 4.5 |
| HfN | 0.9 |

**Table S7: Magnetoresistance of various HfN films at the lowest available temperature.**

| Thickness (nm) | Temp (K) | Magnetoresistance (%) |
|---|---|---|
| 2 | 100 | -0.54 |
| 2.5 | 8.6 | -4.8 |
| 3 | 2.2 | -13.3 |

**Table S8: Experimental carrier concentration ($N_{2D}$), calculated $\Gamma_0$, and theoretical critical carrier concentration ($N_{2D}^C$) for different thicknesses of HfN film.**

| $d\ (nm)$ | $N_{2D}(cm^{-2})$ (Experimental) | $\Gamma_0(n,d)$ | $N_{2D}^C(cm^{-2})$ (Theoretical) |
|---|---|---|---|
| 2 | $7.4 \times 10^{13}$ | 10.88 | $7.6 \times 10^{13}$ |
| 2.5 | $1.9 \times 10^{15}$ | 0.68 | $6.2 \times 10^{13}$ |
| 3 | $3.5 \times 10^{15}$ | 0.35 | $5.2 \times 10^{13}$ |
| 4 | $9.5 \times 10^{15}$ | 0.12 | $3.9 \times 10^{13}$ |
| 5 | $1.2 \times 10^{16}$ | 0.08 | $3.2 \times 10^{13}$ |
| 10 | $3.3 \times 10^{16}$ | 0.02 | $1.6 \times 10^{13}$ |
| 150 | $7.9 \times 10^{17}$ | 0.0001 | $1.1 \times 10^{12}$ |

**Table S9: Calculated EPC constant λ and superconducting critical temperature ($T_c$).**

| Materials | λ | $T_c$ (calculated in K) | $T_c$ (experimental in K) |
|---|---|---|---|
| HfN | 0.72 | 9.15 | 9.18 |
| TiN | 0.56 | 5.65 | 5.35 |
| Au | 0.22 | ----- | ----- |
| Ag | 0.16 | ----- | ----- |

**Table S10: Experimental ENZ wavelength and Drude plasma frequency of HfN films.**

| HfN films | $\lambda_{ENZ}$ (nm) | $\omega_P$ (eV) |
|---|---|---|
| a. Das *et. al.* | 350 | 8.40 |
| b. Das *et. al.* | 370 | 8.31 |
| c. This work | 416 | 7.85 |

**Table S11: Simulated $\lambda_{ENZ}$ and Drude plasma frequency ($\omega_P$) of HfN from SIMPLE code.**

| Tot_charge | $\lambda_{ENZ}$ (nm) | $\omega_P$ (eV) |
|---|---|---|
| 0.000 | 348 | 9.00 |
| 0.023 | 350 | 8.93 |
| 0.035 | 357 | 8.91 |
| 0.117 | 368 | 8.70 |
| 0.234 | 405 | 8.32 |

**Table S12: Thickness-dependent optical properties of HfN calculated from SIMPLE code.**

| Thickness | $\lambda_{ENZ}$ (nm) | $\omega_P$ (eV) |
|---|---|---|
| bulk | 348 | 9.00 |
| 2 nm (10 layers) | 361 | 5.90 |
| 0.2 nm (2 layers) | 450 | 3.54 |

**Supplementary References**

53. Maurya, K. C., Shalaev, V. M., Boltasseva, A. & Saha, B. Reduced optical losses in refractory plasmonic titanium nitride thin films deposited with molecular beam epitaxy. *Opt. Mater. Express* **10**, 2679 (2020).

54. Bondarev, I. V., Mousavi, H. & Shalaev, V. M. Transdimensional epsilon-near-zero modes in planar plasmonic nanostructures. *Phys. Rev. Res.* **2**, 013070 (2020).

55. Liao, J. *et al.* Observation of Anderson Localization in Ultrathin Films of Three-Dimensional Topological Insulators. *Phys. Rev. Lett.* **114**, 216601 (2015).

56. Giannozzi, P. *et al.* Advanced capabilities for materials modelling with Quantum ESPRESSO. (arXiv:1709.10010v1 [cond-mat.mtrl-sci]). *J. Phys. Condens. Matter* **29**, 465901 (2017).

57. Perdew, J. P., Burke, K. & Ernzerhof, M. Generalized Gradient Approximation Made Simple. *Phys. Rev. Lett.* **77**, 3865–3868 (1996).

58. Vanderbilt, D. Soft self-consistent pseudopotentials in a generalized eigenvalue formalism. *Phys. Rev. B* **41**, 7892–7895 (1990).

59. Troullier, N. & Martins, J. L. Efficient pseudopotentials for plane-wave calculations. *Phys. Rev. B* **43**, 1993–2006 (1991).

60. Schlipf, M. & Gygi, F. Optimization algorithm for the generation of ONCV pseudopotentials. *Comput. Phys. Commun.* **196**, 36–44 (2015).

61. Methfessel, M. & Paxton, A. T. High-precision sampling for Brillouin-zone integration in metals. *Phys. Rev. B* **40**, 3616–3621 (1989).

62. Marzari, N., Vanderbilt, D., De Vita, A. & Payne, M. C. Thermal Contraction and

Disordering of the Al(110) Surface. *Phys. Rev. Lett.* **82**, 3296–3299 (1999).

63. Timrov, I. *et al.* Ab initio study of electron energy loss spectra of bulk bismuth up to 100 eV. *Phys. Rev. B* **95**, 094301 (2017).

64. Timrov, I., Vast, N., Gebauer, R. & Baroni, S. Electron energy loss and inelastic x-ray scattering cross sections from time-dependent density-functional perturbation theory. *Phys. Rev. B* **88**, 064301 (2013).

65. Prandini, G., Galante, M., Marzari, N. & Umari, P. SIMPLE code: Optical properties with optimal basis functions. *Comput. Phys. Commun.* **240**, 106–119 (2019).

66. Motornyi, O. *et al.* Electron energy loss spectroscopy of bulk gold with ultrasoft pseudopotentials and the Liouville-Lanczos method. *Phys. Rev. B* **102**, 035156 (2020).

67. Maurilio, H., Carvalho, L., Rebello, M., Dias, S. & Bezerra, A. T. Anisotropic optical response of gold-silver alloys. 0–18 (2021) doi:https://doi.org/10.21203/rs.3.rs-711773/v1.

68. Rangel, T. *et al.* Band structure of gold from many-body perturbation theory. *Phys. Rev. B* **86**, 125125 (2012).

69. Li, S. *et al.* Anomalous thermal transport in metallic transition-metal nitrides originated from strong electron–phonon interactions. *Mater. Today Phys.* **15**, 100256 (2020).

70. Saha, B., Acharya, J., Sands, T. D. & Waghmare, U. V. Electronic structure, phonons, and thermal properties of ScN, ZrN, and HfN: A first-principles study. *J. Appl. Phys.* **107**, (2010).

71. Christensen, A. N., Kress, W., Miura, M. & Lehner, N. Phonon anomalies in transition-metal nitrides: HfN. *Phys. Rev. B* **28**, 977–981 (1983).

72. Gall, D. The search for the most conductive metal for narrow interconnect lines. *J. Appl. Phys.* **127**, 48–52 (2020).

73. Balasubramanian, K., Khare, S. V. & Gall, D. Energetics of point defects in rocksalt structure transition metal nitrides: Thermodynamic reasons for deviations from stoichiometry. *Acta Mater.* **159**, 77–88 (2018).

74. Shah, D. *et al.* Controlling the Plasmonic Properties of Ultrathin TiN Films at the Atomic Level. *ACS Photonics* **5**, 2816–2824 (2018).